\documentclass{aa}  
\usepackage{graphicx}
\usepackage{txfonts}
\usepackage{natbib}
\usepackage{tabularx,booktabs}
\usepackage{cprotect}
\newcommand{\micron}{\,\mu\rm{m}}
\newcommand{\arcs}{^{\prime\prime}}
\usepackage{hyperref}
\hypersetup{colorlinks,linkcolor={blue},citecolor={blue}}

\defcitealias{Beck2022}{B22}

\begin{document}

   \title{Infrared view of the multiphase ISM in NGC\,253}

   \subtitle{II. Modelling the ionised and neutral atomic gas}

   \author{Andr\'{e} Beck
          \inst{1}
          \and
          Vianney Lebouteiller\inst{2}
          \and
          Suzanne C. Madden\inst{2}
          \and
          Aaron Bryant\inst{1}
          \and
          Christian Fischer\inst{1}
          \and
          Christof Iserlohe\inst{1}
          \and
          Maja Ka\'{z}mierczak-Barthel\inst{1}
          \and
          Alfred Krabbe\inst{1}
          \and
          Serina T. Latzko\inst{1}
          \and
          Juan-Pablo P\'{e}rez-Beaupuits\inst{3,4}
          \and
          Lise Ramambason\inst{5}
          \and
          Hans Zinnecker\inst{6}
          }

   \institute{Deutsches SOFIA Institut, University of Stuttgart,
              Pfaffenwaldring 29, D-70569 Stuttgart, Germany\\
              \email{andre.beck1993@gmail.com}
         \and
            AIM, CEA, CNRS, Universit\'{e} Paris-Saclay, Universit\'{e} Paris Diderot, Sorbonne Paris Cit\'{e}, F-91191 Gif-sur-Yvette, France
        \and
            Max-Planck-Institut für Radioastronomie, Auf dem Hügel 69, 53121 Bonn, Germany
        \and
            Centro de Astro-Ingenier\'{i}a (AIUC), Pontificia Universidad Cat\'{o}lica de Chile, Av. Vicuña Mackena 4860, Macul, Santiago, Chile
        \and 
            Institut f\"{u}r Theoretische Astrophysik, Zentrum f\"{u}r Astronomie, Universit\"{a}t Heidelberg, Albert-Ueberle-Str. 2, D-69120 Heidelberg, Germany 
        \and
            Universidad Autonoma de Chile, Pedro de Valdivia 425, Providencia, Santiago de Chile, Chile
			}
   \date{Received xxx / Accepted yyy}

    \abstract
    {
        Multi-wavelength studies of galaxies and galactic nuclei allow us to build a relatively more complete picture of the interstellar medium (ISM), especially in the dusty regions of starburst galaxies. 
        An understanding of the physical processes in nearby galaxies can assist in the study of more distant sources at higher redshifts, which cannot be resolved.
        
    }  
    {
        We aimed to use observations presented in the first part of this series of papers to model the physical conditions of the ISM in the nuclear region of NGC\,253, in order to obtain primary parameters such as gas densities and metallicities.
        From the created model we further calculated secondary parameters such as gas masses of the different phases, and estimated the fraction of $\left[ \ion{C}{ii} \right]158\micron$ from the different phases, which allowed us to probe the nuclear star-formation rate. 
    }
    {
        To compare theory with our observations we used \texttt{MULTIGRIS}, a probabilistic tool that determines probabilities for certain ISM parameters from a grid of \texttt{Cloudy} models together with  a set of spectroscopic lines. 
    }
    {
        We find that the hypothetical active galactic nucleus within NGC\,253 has only a minor impact compared to the starburst on the heating of the ISM as probed by the observed lines.
        We characterise the ISM and obtain parameters such as a solar metallicity, a mean density of $\sim 230\,\mathrm{cm}^{-3}$, an ionisation parameter of $\log U\approx -3$, and an age of the nuclear cluster of $\sim 2\,\mathrm{Myr}$.
        Furthermore, we estimate the masses of the ionised ($3.8\times10^{6}\,M_{\odot}$), neutral atomic ($9.1\times10^{6}\,M_{\odot}$), and molecular ($2.0\times10^{8}\,M_{\odot}$) gas phases as well as the dust mass ($1.8\times10^{6}\,M_{\odot}$) in the nucleus of NGC\,253.
        
    }
    {}


   \maketitle

\section{Introduction}
    In the extreme environments of galactic nuclei, the various heating and cooling mechanisms at work within the interstellar medium (ISM) are notoriously entangled.
    Consequently, the physical conditions of the ISM are difficult to constrain.
    In particular the heating mechanisms are strenuous to unravel, as they are typically and indirectly probed through specific corresponding cooling processes.
    Multi-wavelength observations and models, ideally with tracers sensitive to the different ISM phases, are therefore mandatory to link heating and cooling processes which regulate matter cycle and star formation.
    Nearby galaxies such as NGC\,253, with a distance of $3.5\,\mathrm{Mpc}$ \citep{Rekola2005}, are ideal laboratories to study these effects.
    Our understanding of the physics in nearby galaxies may then be useful for more distant sources, where spatial resolution is far more limited.
    Models of the ISM are becoming ever more complex, able to account for an increasing number of mechanisms, encompassing not only stellar emission, but also the effects of active galactic nuclei (AGN), shock, cosmic rays etc.
    Furthermore, an increasing number of ISM cooling diagnostics can be considered by the models, including many infrared observables, such as atomic fine-structure emission lines.

    In this study we used observations of the nuclear region of NGC\,253, presented in \citet{Beck2022} -- herafter \citetalias{Beck2022} -- from infrared telescopes to model the ionised and neutral atomic gas.
    Although being one of the archetypical starburst galaxies, the major excitation conditions in the centre of this galaxy are not yet well understood \citep[e.g. ][]{Vogler1999,Guenthardt2015}:
    The nuclear starburst likely plays a crucial role in the gas excitation, however, the central supermassive black hole (BH) and associated AGN may have an impact as well \citep{FernandezOntiveros2009}.
    When constraining the impacts of heating on the ISM, it is also necessary to constrain a variety of relevant physical parameters, such as the metallicity, optical depth, or gas density.
    For instance, the latter varies widely within the literature for NGC\,253, depending on the regions observed and the tracers used \citep[e.g.][]{Puche1991,Wall1991,Carral1994,Engelbracht1998}.
    In \citetalias{Beck2022}, we used a subset of the observed emission lines and analytical approach, from which we obtained solar metallicity, an optical depth of $4.35\,\mathrm{mag}$, and a mean density of $\sim 150\,\mathrm{cm}^{-3}$ for the ionised gas within the central $\sim 100\,\mathrm{pc}$. 
    From the more complete picture built by the multi-wavelength study it is also possible to derive simultaneously and self-consistently masses of the ionised and neutral atomic hydrogen as well as the dust mass, which are the goals of this work.
    As of this study, these have only been obtained on much larger scales \citep{Melo2002,Weiss2008}.

    The $\left[ \ion{C}{ii} \right]158\micron$ line has been recently proposed as a powerful probe to determine not only the local star-formation rate \citep[SFR,][]{HerreraCamus2018}, but also for the CO-dark gas \citep{Madden2020}.
    However, the validity of $\left[ \ion{C}{ii} \right]158\micron$ as a tracer is hindered by its ubiquitous nature, potentially originating from ionised, atomic and molecular phases of the ISM.    
    An objective in this study is to determine masses of the different phases, and calculate the origin of the $\left[ \ion{C}{ii} \right]158\micron$ from the different phases.
    Furthermore, from our models we estimate and compare SFRs from different tracers.
    
    This paper is organised as follows:
    In Sect. \ref{sec:Modellingapproach} we explain the model grid and the code used to infer ISM parameters from this grid.
    Section \ref{sec:modelIonised} continues with creating a model for the ionised gas, by using emission lines that originate primarily in \ion{H}{ii} regions.
    This model serves as a starting point to create a model for the ionised and neutral atomic gas in Sect. \ref{sec:neutralModel}.
    The discussion in Sect. \ref{sec:discussion} presents parameters that can be inferred from the model explained in Sect. \ref{sec:neutralModel}.


\section{Modelling approach}\label{sec:Modellingapproach}
\subsection{Observational constraints}
\begin{figure}
        \centering
        \resizebox{\hsize}{!}{\includegraphics[scale=1]{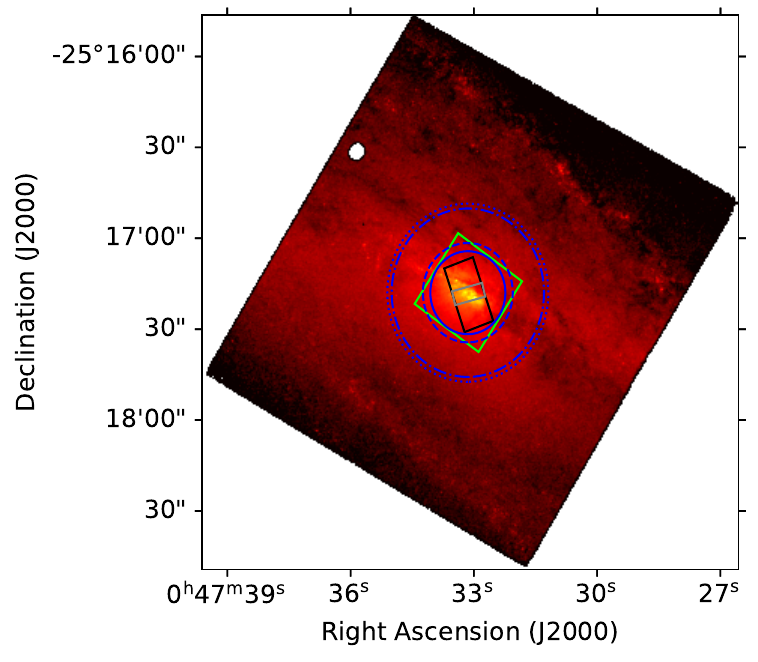}}
        \caption{Optical image from \textit{Hubble}/WFC3 observations of the central $\sim 2^{\prime}$ of NGC\,253. Overlays show the apertures used to extract the line fluxes from the different observatories. The apertures from \textit{Spitzer}/IRS are grey (short-high) and black (long-high), green shows the footprint of the \textit{Herschel}/PACS integral-field-unit. Blue circles are the apertures for the SOFIA/FIFI-LS observations of $\left[ \ion{O}{iii} \right]52\micron$ (solid), $\left[ \ion{O}{iii} \right]88\micron$ (dashed), $\left[ \ion{O}{i} \right]146\micron$ (dashdot), and $\left[ \ion{C}{ii} \right]158\micron$ (dotted) as described in \citetalias{Beck2022}. See also Table \ref{tab:ionisedSystematicOffsets}.}
        \label{fig:aperturesPaperI}
    \end{figure}

    In \citetalias{Beck2022} we presented far-infrared observations from SOFIA/FIFI-LS, complemented by mid-infrared and far-infrared archival data from \textit{Herschel}/PACS, \textit{Herschel}/SPIRE, and \textit{Spitzer}/IRS.
    Apertures are shown in Fig. \ref{fig:aperturesPaperI}.
    We obtained line fluxes and corresponding line flux errors of $30$ emission lines, covering different species (C, N, Ne, O, Si, \dots) and ionisation states of the nuclear region of NGC\,253.
    The observed emission lines are sensitive to various density regimes, due to their wide range of critical densities.
    For instance the critical densities for $\left[\ion{Ne}{ii}\right]13\micron$ and $\left[\ion{N}{ii}\right]205\micron$ are $7\times 10^{5}\,\mathrm{cm}^{-3}$ and $45\,\mathrm{cm}^{-3}$, respectively.
    The different ionisation states with ionisation potentials up to $97.12\,\mathrm{eV}$ to create Ne$^{4+}$ also show the variety of physical conditions in the centre of NGC\,253.
    Owing to the large range of ionisation potentials and density regimes to excite these infrared emission lines, we are now able to draw a more complete picture of the embedded regions in the nucleus of NGC\,253 and estimate the physical conditions there, by modelling all lines simultaneously. 

    In addition to the line fluxes listed in Table 3 of \citetalias{Beck2022} we also used observations of the CO rotational transitions CO$\left(1-0\right)$ to CO$\left(3-2\right)$ from \citet{Israel1995} and CO$\left(4-3\right)$ to CO$\left(8-7\right)$ from \citet{PerezBeaupuits2018} to compare \textit{a posteriori}.
    Higher-$J$ CO emission lines are also available in \citet{PerezBeaupuits2018}.
    However, these lines are not available in the model grid (Sect. \ref{ssec:modelGrid}) and hence we could not use them as constraints nor for a comparison \textit{a posteriori}.

\subsection{Model grid}\label{ssec:modelGrid}
    The grid of models used in this study is \textit{Star-Forming Galaxy with an X-ray component} \citep[SFGX,][]{Ramambason2022}\footnote{SFGX is available within the \texttt{MULTIGRIS} framework at \url{https://gitlab.com/multigris/mgris}}.
    SFGX is a grid of \texttt{Cloudy} \citep[a 1D radiative transfer code, see][]{Ferland2017} models assuming a cloud of spherical geometry.
    The modelled cloud is illuminated by a representative stellar cluster of a given age $t$ and luminosity $L^{\star}$.
    The spectrum of this star cluster is created with BPASS \citep{Stanway2018}.
    In addition to the star cluster, an X-ray source with inner disk temperature $T_{X}$\footnote{The outer disk temperature is assumed to be a constant $T_{X,\,\mathrm{out}} = 10^{3}\,\mathrm{K}$.} and luminosity $L_{X}$ is modelled by a multi-colour black body \citep{Mitsuda1984}.
    In this study we refer to this combination of star cluster and X-ray source, which are assumed to be co-spatial, as \textit{cluster}.
    A cluster in this context is described by a parameter set $(L^{\star},\, t,\, L_{X},\, T_{X})$.
    The X-ray source in these \texttt{Cloudy} models is particularly important in our study, because of the presence of highly ionised emission lines such as $\left[\ion{Ne}{v}\right]14\micron$ and $\left[\ion{O}{iv}\right]26\micron$.
    Ions like Ne$^{4+}$ and O$^{3+}$ with ionisation potentials $> 54.9\,\mathrm{eV}$ (or $\lambda = 22.6\,\mathrm{nm}$) to create the ion, are unlikely to be created by stars and likely imply the presence of an AGN or a different source of X-ray emission.

    An ISM component in the grid is characterised by the metallicity $Z$, ionisation parameter $U$, hydrogen volume density at the illuminated edge of the cloud $n$, and the depth.
    We emphasise that the metallicity refers to the O/H abundance.
    Other elements (such as N, Ne, C, Ne, Si) scale with the O/H abundance -- see \citet{Ramambason2022} for the respective relations.
    Since the luminosity is already included in the cluster properties, the ionisation parameter is defined by the distance $r$ between the cloud and the cluster.
    SFGX was created to investigate physical conditions in low-metallicity galaxies.
    Since the  optical depth $A_{V}$ as a function of depth varies with metallicity (i.e. two clouds with the same physical size but different metallicities have different $A_{V}$), a more general definition of the cloud depth was defined by a ``cut'' parameter $\xi$ as shown in Table \ref{tab:cutParamDefinition}.

    \begin{table}
        \centering
        \caption{Definition of the different values for the cut parameter $\xi$ in the SFGX grid. There are tabulated decimal values between the $\xi$ values.}
        \begin{tabular}{lll}
            \toprule
            \toprule
            $\xi$ & Condition & Note \\
            \midrule
            $0$ &  & illuminated edge of the cloud \\
            $1$ & $n(\mathrm{H^{+}})/n(\mathrm{H^{0}}) = 1.0$ & ionisation front \\
            $2$ & $n(\mathrm{H^{0}})/n(\mathrm{H_{2}}) = 1.0$ & hydrogen photodissociation front \\
            $3$ & $n(\mathrm{C^{0}})/n(\mathrm{CO}) = 1.0$ & CO photodissociation front\tablefootmark{a} \\
            $4$ & $A_{V} = 10$ & full depth \\
            \bottomrule
        \end{tabular}
        \tablefoot{
            \tablefoottext{a}{In some models ---in particular ones with large X-ray luminosities or high CRIR that keep the ISM warm--- the CO photodissociation front is not reached within the \texttt{Cloudy} simulation.}
            }
        \label{tab:cutParamDefinition}
    \end{table}
    
    In the following a parameter set ($Z$, $U$, $n$, $\xi$) will be called a \textit{component}.
    Note that in case of modelling more than one component, the metallicity is the same for all components.
    The SFGX grid contains a total of $32000$ models, which are divided into $544000$ sub-models by introducing $17$ cut values between $\xi=0$ and $\xi=4$.
    We refer to \citet{Ramambason2022} for more details about the grid of models and an application to the \textit{Herschel} Dwarf Galaxy Survey \citep{Madden2013}.

\subsection{Inference method}
    We used \texttt{MULTIGRIS} for the modelling approach \citep{Lebouteiller2022}.
    \texttt{MULTIGRIS} uses a Bayesian approach, designed to calculate posterior probability density distributions of parameters of a given grid of models, by using priors (e.g. line fluxes and errors) as constraints.
    Upper and lower limits for line fluxes can be used as well.

    \texttt{MULTIGRIS} allows a linear combination of components, or continuous distributions for the parameters.
    This is particularly important, as the ISM in galaxies is not homogeneous on \mbox{(kilo-)}parsec-scales, but instead heterogeneous and a mix of several components \citep[e.g.][]{Lacy1982,Burton1990,Snow2006,Cormier2019}.
    By considering several components or using, for instance, power-laws for certain parameters, various geometries that are relevant to studies of complex objects like galaxies can be explored.
    Hence, we investigated two different model architectures:
    a one cluster and two component architecture (1C2S), and an architecture where $U$, $n$, $\xi$, and $t$ are continuously distributed following a power-law (PLaw).
    Both architectures have been used to model the ISM in galaxies, see for instance \citet{Pequignot2008,Polles2019} approaches with a discrete sampling and \citet{Baldwind1995,Richardson2016,Ramambason2023} for continuous distribution of ISM parameters.
    A power-law approach, for instance for the depth and density takes the porosity and clumpiness of the ISM into account, by enabling a larger number of low density (diffuse) clouds and only a few clouds of higher density and large depth, or vice versa.
    Each parameter that is described by a power-law introduces three variables: 
    the slope $\alpha$ of the power-law, and the lower and upper boundary, between which the power-law is valid.
    
    The choice for a probabilistic approach is justified by the complexity of the problem, for example:
    \begin{itemize}
        \item The combination of two or more components (i.e. stellar populations or gas phases) yields a large number of free parameters making it difficult to gauge degenerate solutions.
        \item Upper limits or asymmetric uncertainties of line fluxes are difficult to handle with deterministic approaches.
        \item Known parameters (e.g. the metallicity) and their uncertainty cannot be used as priors in a deterministic approach.
    \end{itemize}
    Probabilistic tools such as \texttt{MULTIGRIS} are able to overcome all the mentioned issues of deterministic techniques.
    For more details about the methodology, sampling techniques and applications of \texttt{MULTIGRIS} see \citet{Lebouteiller2022}. 
       
    We note that in \citetalias{Beck2022} we listed the obtained line fluxes and statistical uncertainties in Table 3.
    The uncertainties did not include systematical or calibration errors.
    We included these uncertainties by defining line sets for each instrument.
    Each line set includes the emission lines observed by one instrument.
    We assume that systematic and calibration uncertainties affect the observed emission lines of an instrument all in the same way.
    For each line set, we introduced a scaling parameter to account for these systematic and calibration uncertainties.
    This also allowed us to consider potential offsets due to the different spatial resolution and wavelength coverage and hence the different size of the nucleus.
    In particular we also divided the \textit{Spitzer}/IRS emission lines in two observation sets since this instrument consists of two different modules with different aperture sizes, namely short-high and long-high modules.
    See the discussion in Sect. 2.3.2 of \citetalias{Beck2022} for details.    
    
    

\section{Model for the ionised gas}\label{sec:modelIonised}
    In this first step we investigated model results using the emission lines arising from \ion{H}{ii} regions only, meaning all lines with an ionisation potential $\geqslant 13.6\,\mathrm{eV}$.
    We note that lines with a lower ionisation potential (for example $\left[ \ion{C}{ii} \right]158\micron$ and $\left[ \ion{Si}{ii} \right]35\micron$) may also arise in \ion{H}{ii} regions \citep[e.g.][]{Abel2005,Chevance2016}.
    However, we did not use these lines as constraints in the first step, because of their ambiguous origin.
    But we let the model predict which fraction of $\left[ \ion{C}{ii} \right]158\micron$ comes from the ionised gas only.
    This is done by comparing the cumulative line flux of $\left[ \ion{C}{ii} \right]158\micron$ at the ionisation front and at the solution for the cut-value $\xi$.
    Lines with lower ionisation potentials will be included in Sect. \ref{sec:neutralModel}.
    Also we note that the $\left[ \ion{Fe}{iii} \right]23\micron$ emission line is not included -- due to the strong dependence of Fe emission lines on the dust-to-gas ratio, these emission lines will be handled separately (see Sect. \ref{ssec:results_neutral}).
    
    Recently, \citet{Behrens2022} reported that the cosmic ray ionisation rate (CRIR) in selected clouds in the nuclear region of NGC\,253 is about $10^{4}\times$ larger than the average Galactic value $\zeta_{0} = 2\times 10^{-16}\,\mathrm{s}^{-1}$ \citep{Indriolo2007}. 
    The value for the CRIR used in SFGX, however, is only $3\zeta_{0}$ \citep{Ramambason2022}.
    We will show in Sect. \ref{ssec:ionisedModelCRIReffect} that cosmic rays only mildly affect the ionised gas and hence the effect on the model results are negligible.
    The effect of the increased CRIR on the neutral atomic and molecular gas, however, is significant and will be explored on a smaller sub-grid (see Sect. \ref{sec:neutralModel}).

\subsection{Systematic offsets within the dataset}\label{ssec:ionised_offsets}
    In \citetalias{Beck2022} we showed the data reduction and cross-calibration of observations of NGC\,253 with SOFIA, \textit{Spitzer}, and \textit{Herschel}.
    To account for potential offsets between the line sets (i.e. the emission lines as observed by one instrument) due to calibration or aperture effects, we performed \texttt{MULTIGRIS} runs where the different line sets were successively added (i.e. first run with IRS short-high (SH) and long-high (LH), second run adding the PACS lines etc.).
    \texttt{MULTIGRIS} determines the -- small -- systematic offsets between datasets so that observations match better with models.
    The LH module of \textit{Spitzer}/IRS served as standard (i.e. offset$_{\mathrm{LH}}\equiv 1$).
    As the aperture size of this instrument matches the best with the angular size of the nucleus ($6.68\arcs$), we believe that this observation contains low background emission and we therefore expect the LH line set to be of the best quality.
    In this way, systematic offsets between line sets should be the dominant source of uncertainties.
    To simultaneously investigate the effect of different model architectures, we performed the same runs for a 1C2S and for a PLaw architecture.
    These architectures fulfil the request for more than one component as claimed in \citetalias{Beck2022}, but are still as simple as possible.
    Due to the limited number of constraints, a simple model with fewer parameters preferably avoids overfitting.
    Table \ref{tab:ionisedSystematicOffsets} shows the obtained offsets between observation sets that will be further used in this work.
    Line fluxes shown in Table 3 of \citetalias{Beck2022} were multiplied by these factors to account for the offsets introduced by calibration uncertainties and PSF effects between the instruments.

    \begin{table}
        \centering
        \caption{Mean systematic line flux offsets due to calibration or aperture effects between observation blocks determined from \texttt{MULTIGRIS} runs in linear scale. Only ionised gas lines were used as constraints. The aperture sizes (in $\arcs$ and pc) are added for clarity.}
        \begin{tabular}{lllcc}
            \toprule
            \toprule
            Instrument & 1C2S & PLaw & \multicolumn{2}{c}{Aperture size} \\
                       &      &      & [$\arcs$] & [pc] \\
            \midrule
            IRS/LH\tablefootmark{a}  & $1.0$ & $1.0$  & $11.1 \times 22.3$ & $189 \times 379$\\
            IRS/SH\tablefootmark{a}  & $0.6$ & $0.9$  & $4.7 \times 11.3$  & $80 \times 192$\\
            PACS\tablefootmark{a}    & $0.5$ & $0.5$  & $28.2 \times 28.2$ & $479 \times 479$\\
            FIFI-LS\tablefootmark{b} & $0.6$ & $0.5$  & $13.7 - 29.4$      & $233 - 500$\\
            SPIRE\tablefootmark{b}   & $0.6$ & $0.6$  & $20.0$             & $340$\\
            \bottomrule
        \end{tabular}
        \label{tab:ionisedSystematicOffsets}
        \tablefoot{
            \tablefoottext{a}{Edge length of the rectangular aperture (Fig. \ref{fig:aperturesPaperI}).}\\
            \tablefoottext{b}{Radius of the circular aperture. See \citetalias{Beck2022} for the exact SOFIA apertures and their determination.}
            For comparison, the effective radius of the nucleus in the near-infrared  bulge is $9.1\arcs$ or $150\,\mathrm{pc}$ \citep{Iodice2014}.
            }
    \end{table}

    The systematic offsets of all line sets are smaller than $1$.
    This could be due to different beam sizes, meaning that the \textit{Herschel} and SOFIA observations include more background emission because of their lower spatial resolution.
    This would also explain that the PACS, SPIRE, and FIFI-LS offsets are all in good agreement.
    The short-high observations of \textit{Spitzer}/IRS are also $< 1$.
    Since this observation does not fully sample the nucleus due to a small aperture size, we performed a scaling correction of $1.55$ based on the continua of SH and LH in \citetalias{Beck2022}.
    The lines may not arise from the exact same region as the thermal emission at $\sim 20\micron$, which could lead to an over-correction of the line fluxes and thus a systematic offset $<1$ in the SH emission lines.
    
\subsection{Impacts of the model configuration}\label{ssec:ionisedModelConfig}
    
    To evaluate the performance, we used the marginal likelihood as well as ``$pn\sigma$'' values, both of which are computed by \texttt{MULTIGRIS}.
    The marginal likelihood measures the probability that the model grid including any prior reproduces the observations and will reach a maximum for the most likely set of parameters under the assumption that the priors do not change.
    Too few parameters are not able to reproduce all the emission lines and will therefore have a low marginal likelihood, while too many parameters will cause overfitting, that is that the model contains more parameters than can be justified by the constraints.
    The $pn\sigma$ values on the other hand describe how many draws of the posterior distribution agree within $n\sigma$ of the constraints or observables.
    The higher the $pn\sigma$ values, the better the model is able to reproduce the observations within their uncertainties.
   
\begin{figure*}
        \centering
        \includegraphics[width=17cm]{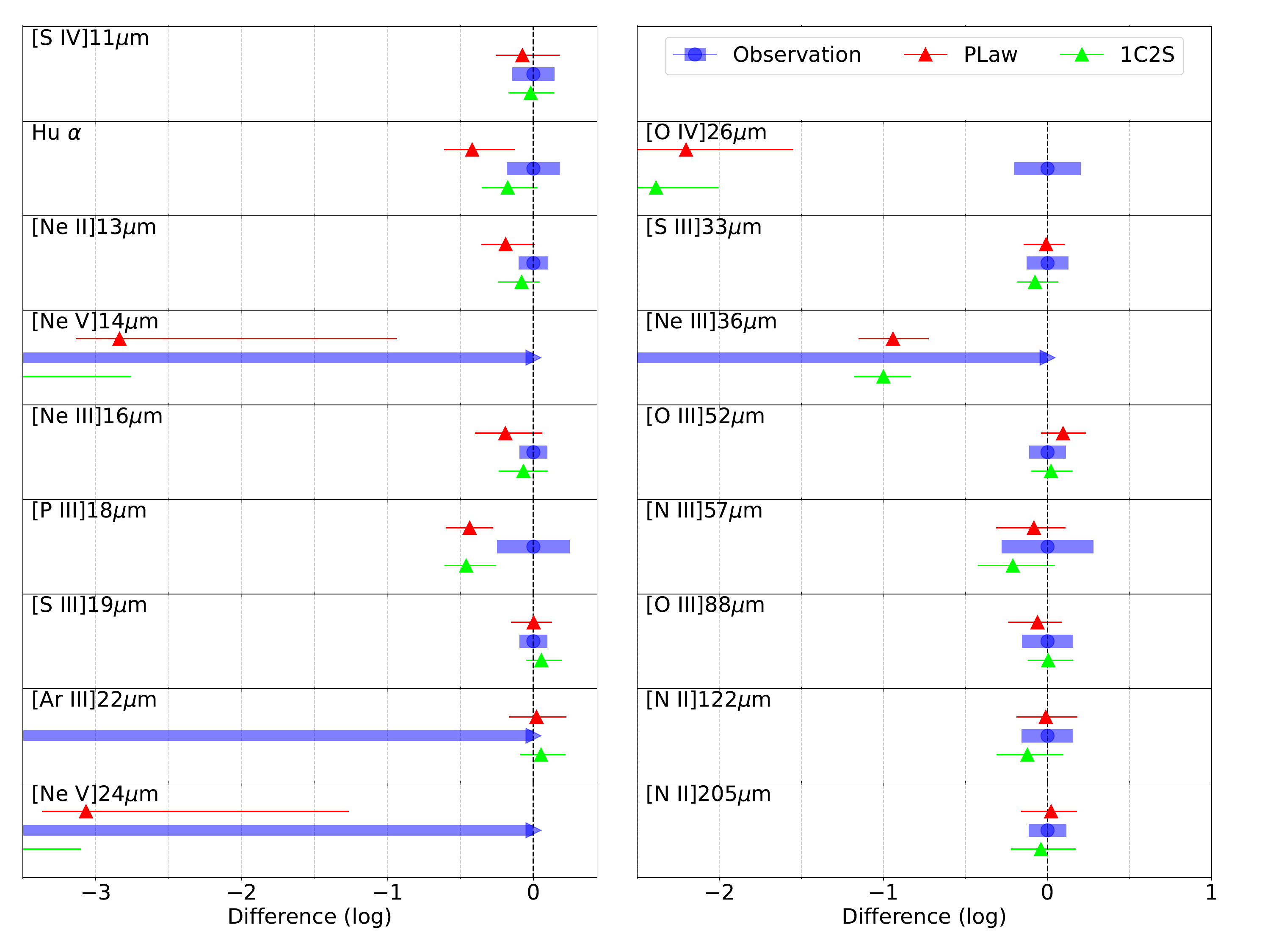}
        \caption{Comparison of modelled/predicted and observed line fluxes. Emission lines from the ionised gas are used as constraints. The abscissa is normalised to the observed line flux ($\equiv 0$) with errorbars (blue). Red and green show the resulting line fluxes from the PLaw and 1C2S architecture respectively. For display reasons, the abscissa in column one are compressed so that the PLaw results for the $\left[ \ion{Ne}{v} \right]$ lines are not entirely visible. Blue arrows denote upper limits on line fluxes.}
        \label{fig:ionised_lineFluxComp_PLaw_1C2S}
    \end{figure*}

    Generally, both architectures show high $pn\sigma$ values with $p3\sigma = 99.19\%$ and $p3\sigma = 96.44\%$ for the 1C2S and PLaw architecture, respectively.
    Moreover, both are able to reproduce all emission lines (with the exception of $\left[ \ion{O}{iv} \right]26\micron$) within the uncertainties (see Fig. \ref{fig:ionised_lineFluxComp_PLaw_1C2S}).
    The log marginal likelihood is better for the 1C2S architecture ($-31.8$ compared to $-37.3$ for the PLaw architecture).
    However, this does not mean that the 1C2S architecture is a more realistic or ``better'' one -- see for instance \citet{Lebouteiller2023} for a discussion on model architectures.
    To conclude, the model architecture does not seem to play a significant role for the modelling of the ionised gas.
    
\subsection{Results for the ionised gas}\label{ssec:ionisedResults}
\begin{figure}
        \centering
        \resizebox{\hsize}{!}{\includegraphics[scale=1]{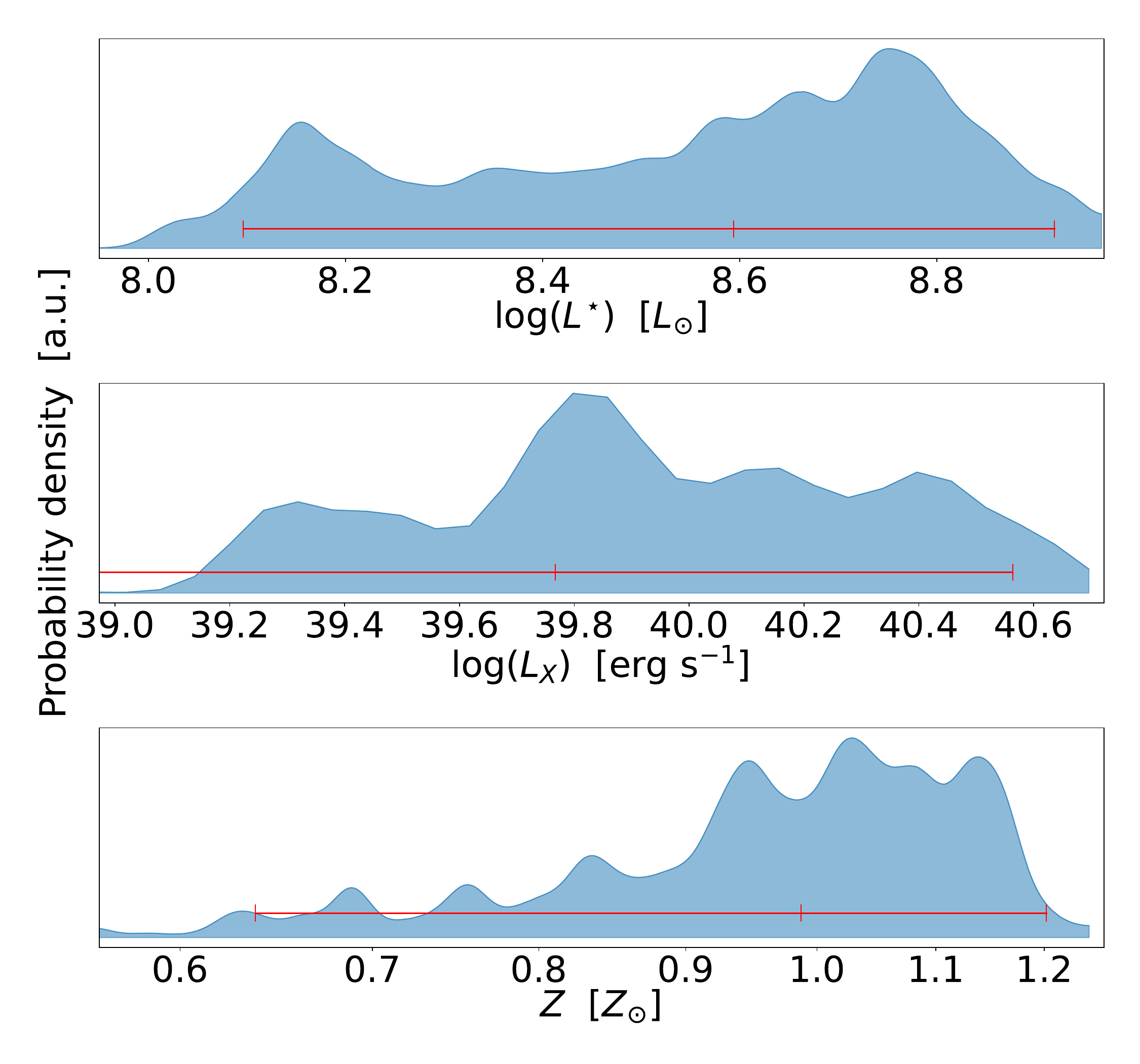}}
        \caption{Probability density distributions for the stellar luminosity $L^{\star}$, X-ray luminosity $L_{X}$, and metallicity $Z$ for the PLaw architecture (see Sect. \ref{ssec:ionisedResults}). Red shows the median and confidence intervals of the posterior distribution.}
        \label{fig:traceParamsIonised_LLXZ}
    \end{figure}

    We started by narrowing down the parameter space of our model grid.
    This is necessary since a new model grid with a larger CRIR is required but the full set of \texttt{Cloudy} models would take a long time to run.
    To do so, we investigated parameters that can already be determined from the solution for the ionised gas.
    For instance, emission lines from species with high ionisation potentials such as $\left[ \ion{O}{iv} \right]26\micron$ ($55\,\mathrm{eV}$) and $\left[ \ion{Ne}{v} \right]14\micron$ ($97\,\mathrm{eV}$) are almost exclusively created by X-rays \citep{Abel2008}.
    Hence, the parameters of the X-ray source are already constrained when using these emission lines.
    Another parameter that can already be determined is the metallicity $Z$, provided we assume that the metallicity does not vary significantly within the nuclear region.
    We further estimate the stellar luminosity $L^{\star}$ from the ionised gas model, assuming that this parameter should not vary greatly once the neutral atomic gas lines are taken into account.
    We focus on the PLaw architecture in this section and will compare the results with those from the 1C2S architecture at the end of this section.
    Figure \ref{fig:traceParamsIonised_LLXZ} shows the probability density functions (PDF) of the parameters determined in this section.
    \citet{MejiaNarvaez2020} interpret the PDFs, for instance for the metallicity, as a physical distribution within the ISM, however, we think that the PDF in our case is dominated by the uncertainty rather than a physical metallicity distribution function.  
    Table \ref{tab:ionisedParamsResults} shows the resulting median and standard deviations for all variables of the model.
    For the power-law distributed parameters an average and standard deviation is given as well.
    

    \begin{table*}
        \centering
        \caption{Resulting mean values and confidence intervals from the inference for the PLaw model from SFGX (see Sect. \ref{ssec:ionisedResults}) and SFNX (see Sect. \ref{sec:neutralModel}). Both models results from runs where only emission lines from the ionised gas are considered as constraints.}
        \begin{tabular}{lrrrr}
            \toprule
            \toprule
            Parameter                         & SFGX (PLaw)             & SFGX (1C2S)             & SFNX (PLaw)                 & SFNX (1C2S) \\
            \midrule
            $L^{\star}$ [$10^{8}\,L_{\odot}$] & $7.70_{-5.25}^{+8.58}$  & $6.00_{-3.84}^{+9.40}$  & $10.0$\tablefootmark{a}     & $10.0$\tablefootmark{a} \\
            $Z$ [$Z_{\odot}$]                 & $0.99_{-0.35}^{+0.21}$  & $0.71_{-0.26}^{+0.52}$  & $1.0$\tablefootmark{a}      & $1.0$\tablefootmark{a} \\
            $L_{X}$ [$10^{39}\,$erg s$^{-1}$] & $5.9_{-5.9}^{+31.03}$   & $4.59_{-4.59}^{+35.97}$ & $0.01_{-0.01}^{+0.44}$      & $0.01_{-0.01}^{+1.95}$ \\
            $T_{X}$ [$10^{5}\,$K]             & $3.0_{-2.97}^{+86.91}$  & $2.13_{-2.13}^{+89.81}$ & $6.04_{-6.03}^{+81.68}$     & $5.33_{-5.33}^{+72.17}$ \\
            $\zeta$ [$10^{-16}\,$s$^{-1}$]    & $6.0$\tablefootmark{a}  & $6.0$\tablefootmark{a}  & $102.81_{-89.33}^{+635.33}$ & $30.25_{-23.28}^{+362.88}$ \\
            \midrule
            $\alpha_{t}$ [1]                  & $-0.49_{-3.26}^{+2.81}$ &                         & $-0.87_{-3.32}^{+2.6}$      &  \\
            $t_{\mathrm{lower}}$ [Myr]        & $1.59_{-0.53}^{+0.94}$  &                         & $1.63_{-0.55}^{+1.47}$      &  \\
            $t_{\mathrm{upper}}$ [Myr]        & $3.45_{-2.0}^{+2.62}$   &                         & $4.82_{-2.6}^{+4.08}$       &  \\
            $\bar{t}$ [Myr]                   & $2.54_{-1.01}^{+1.04}$  & $3.02_{-1.43}^{+2.57}$  & $2.92_{-1.26}^{+2.46}$      & $3.34_{-1.59}^{+3.21}$ \\
            \midrule
            $\alpha_{n}$ [1]                  & $-1.4_{-3.31}^{+2.39}$  &                         & $-1.16_{-3.32}^{+2.82}$     &  \\
            $n_{\mathrm{lower}}$ [cm$^{-3}$]  & $40_{-36}^{+253}$       &                         & $20_{-19}^{+158}$           &  \\
            $n_{\mathrm{upper}}$ [cm$^{-3}$]  & $562_{-529}^{+5106}$    &                         & $333_{-287}^{+5760}$        &  \\
            $\bar{n}$ [cm$^{-3}$]             & $135_{-121}^{+622}$     & $19_{-18}^{+612}$       & $103_{-81}^{+362}$          & $16_{-14}^{+260}$ \\
            \midrule
            $\alpha_{U}$ [1]                  & $-1.21_{-2.2}^{+3.41}$  &                         & $-1.9_{-0.67}^{+0.74}$      &  \\
            $\log U_{\mathrm{lower}}$ [1]     & $-3.4_{-0.46}^{+0.79}$  &                         & $-3.74_{-0.18}^{+0.23}$     &  \\
            $\log U_{\mathrm{upper}}$ [1]     & $-2.59_{-0.75}^{+2.08}$ &                         & $-2.55_{-0.86}^{+1.09}$     &  \\
            $\log \bar{U}$ [1]                & $-3.0_{-0.44}^{+0.64}$  & $-3.4_{-0.58}^{+0.80}$  & $-3.25_{-0.35}^{+0.3}$      & $-3.4_{-0.50}^{+0.83}$ \\
            \midrule
            $\alpha_{\xi}$ [1]                & $-0.48_{-2.62}^{+2.79}$ &                         & $-0.91_{-2.55}^{+2.25}$     &  \\
            $\xi_{\mathrm{lower}}$ [1]        & $0.65_{-0.51}^{+1.2}$   &                         & $0.72_{-0.57}^{+0.7}$       &  \\
            $\xi_{\mathrm{upper}}$ [1]        & $2.06_{-1.02}^{+1.35}$  &                         & $1.85_{-0.97}^{+0.89}$      &  \\
            $\bar{\xi}$ [1]                   & $1.55_{-0.78}^{+0.90}$  & $1.01_{-0.78}^{+0.93}$  & $1.39_{-0.7}^{+0.66}$       & $1.17_{-0.79}^{+0.79}$ \\
            $W_{1}$ [1]                       & $0.58^{+0.38}_{-0.53}$  &                         & $0.68^{+0.24}_{-0.62}$      &  \\
            $W_{2}$ [1]                       & $0.42^{+0.53}_{-0.38}$  &                         & $0.32^{+0.62}_{-0.24}$      &  \\            
            \bottomrule
        \end{tabular}
        \tablefoot{
            \tablefoottext{a}{Fixed in the respective grid.}
        }
        \label{tab:ionisedParamsResults}
    \end{table*}

    Table \ref{tab:ionisedParamsResults} and Fig. \ref{fig:traceParamsIonised_LLXZ} show that the inferred metallicity from the inference is around the solar value.
    This is in good agreement with results from our analytic approach in \citetalias{Beck2022}, where we calculated $Z=1.0\,Z_{\odot}$ using the $(\left[ \ion{Ne}{ii} \right]13\micron + \left[ \ion{Ne}{iii} \right]16\micron) / \mathrm{Hu}\, \alpha$ line flux ratio.
    The probability is little outside the given uncertainties.
    
    With $7.70_{-5.25}^{+8.58} \times 10^{8}\,\mathrm{L}_{\odot}$ we obtained a somewhat lower stellar luminosity $L^{\star}$ than previous studies, although with a large uncertainty and broad PDF as can be seen in row $1$ of Fig. \ref{fig:traceParamsIonised_LLXZ}.
    \citet{Watson1996} estimated a luminosity from young stars of $1.5\times 10^{9}\,\mathrm{L}_{\odot}$.
    \citet{Radovich2001} calculated $1.5\times10^{10}\,\mathrm{L}_{\odot}$, although with a much poorer spatial resolution ($\sim2^{\prime}$) and a higher extinction correction of $11\,\mathrm{mag}$, compared to our result of $A_{V} = 4.35\,\mathrm{mag}$ in \citetalias{Beck2022}.
    
    The mean age of the nuclear stellar cluster that we obtained from our solution is $2.54_{-1.01}^{+1.04}\,\mathrm{Myr}$.
    This is slightly younger than previous estimates such as \citet{Kornei2009, FernandezOntiveros2009} who obtained $5.7$ and $6.3\,\mathrm{Myr}$, respectively, perhaps due to a different choice of initial-mass-functions (IMF).
    We will investigate effects of different IMFs in Sect. \ref{sec:effect_IMF}.

    One major question that is tackled with this study is the characterisation of the central BH, because the physical properties and the impact of the BH on the ISM are not completely understood.
    The total X-ray luminosity obtained from our models, which we assume to originate from the central BH or AGN is $5.9 \times 10^{39}\,\mathrm{erg\,s}^{-1}$, which is in good agreement with Chandra observations \citep[$4.7\times 10^{39}\,\mathrm{erg\,s}^{-1}$,][]{Lopez2023} and \citet{FernandezOntiveros2009}.
    The X-ray luminosity shows large uncertainties, but the PDF shows only low probabilities for luminosities below $10^{39}\,\mathrm{erg\,s}^{-1}$ and above $4\times10^{40}\,\mathrm{erg\,s}^{-1}$ (see in Fig. \ref{fig:traceParamsIonised_LLXZ}).
    To confirm that an X-ray component is important in this model, we made runs with no X-ray source.
    For both architectures, the $pn\sigma$ values and marginal likelihood drop by several percents, and the highly ionised emission lines are not well reproduced.
    The X-ray component is indeed necessary to model the observed emission lines, in particular to recover the $\left[ \ion{O}{iv} \right]$ and $\left[ \ion{Ne}{v} \right]$ emission.
    For comparison, Sgr\,A$^{*}$ has a similar luminosity of $L_{X} = 4 \times 10^{39}\,\mathrm{erg\,s}^{-1}$ \citep{Kaneda1997}, suggesting that the central BH of NGC\,253 resembles a low-luminosity AGN rather than a typical extragalactic AGN which have luminosities in the range of  $L_{X} = 10^{40} - 10^{43}\,\mathrm{erg\,s}^{-1}$ \citep[e.g.][]{Fornasini2018}.
 
    The power-law distributions of the density $n$, ionisation parameter $U$, and cut parameter $\xi$ show that a model with low density (i.e. diffuse), low depth, and low ionisation parameter clouds is preferred.
    However, the range of these three parameters is small (see lower and upper boundaries in Table \ref{tab:ionisedParamsResults}), suggesting that most of the ionised gas emission arises from a mostly diffuse gas component.
    The average gas density in the power-law model is $135\,\mathrm{cm}^{-3}$, which is in good agreement (although with large uncertainties) with our analytic results in \citetalias{Beck2022}, where we used the $\left[ \ion{O}{iii} \right]52/88\micron$, $\left[ \ion{S}{iii} \right]19/33\micron$, and $\left[ \ion{N}{ii} \right]122/205\micron$ line flux ratios and obtained $84 \lesssim n \lesssim 212\,\mathrm{cm}^{-3}$.
    As expected, the range for cut-values $\xi$ is somewhat beyond the ionisation front $\xi= 1.0$.
    A fraction of the emission from species with ionisation potentials near $13.6\,\mathrm{eV}$ (e.g.$\left[ \ion{N}{ii}\right]$) possibly originates beyond the ionisation front, which is why $\xi \equiv 1.0$ or below is not an expected or reasonable solution.

    In Table \ref{tab:ionisedParamsResults} we also show the resulting parameters for the 1C2S configurations.
    Both, the PLaw and 1C2S architecture are in good agreement within the uncertainties, showing again that the choice of the configurations seems to be only a second order effect when considering only the ionised gas emission lines.
    
\subsection{Influence of cosmic rays}\label{ssec:ionisedModelCRIReffect}
\begin{figure*}
        \centering
        \includegraphics[width=17cm]{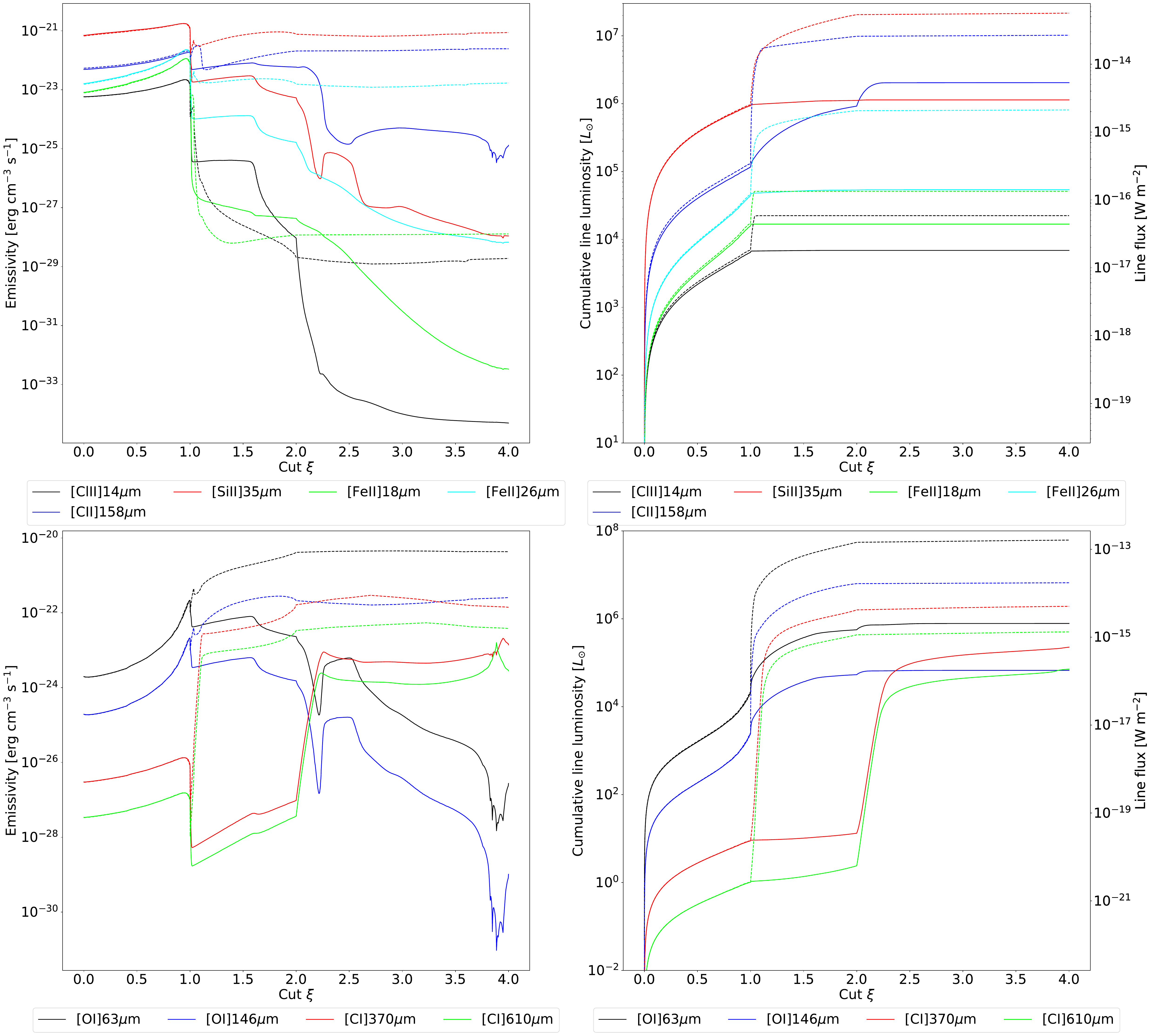}
        \caption{Emissivity (\textit{left}) and cumulative line flux (\textit{right}) over the cut parameter $\xi$ for emission lines from the ionised gas. Solid lines show models with a low CRIR ($6\times10^{-16}\,\mathrm{s}^{-1}$), dashed lines are from models with a high CRIR ($10^{-13}\,\mathrm{s}^{-1}$).}
        \label{fig:CRIR_comparison_neutral}
    \end{figure*}

\begin{figure*}
        \centering
        \includegraphics[width=17cm]{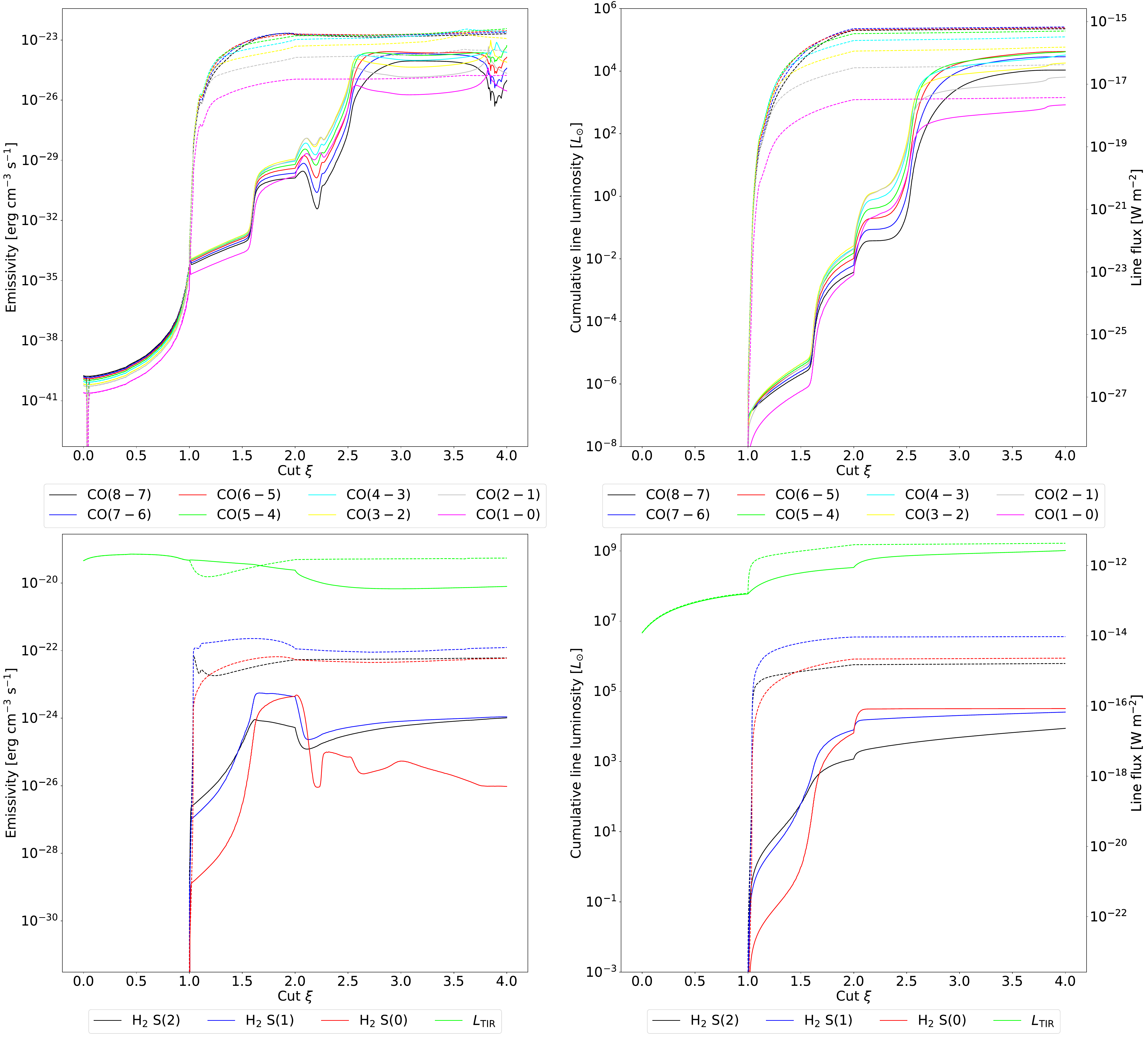}
        \caption{Same as Fig. \ref{fig:CRIR_comparison_neutral} but for emission lines from the molecular gas and the total infrared luminosity $L_{\mathrm{TIR}}$.}
        \label{fig:CRIR_comparison_mol}
    \end{figure*}

    Using data from the ALCHEMI spectral survey \citep{Martin2021}, \citet{Behrens2022} showed that the CRIR in the centre is three to four orders of magnitudes larger than the average Milky Way value $\zeta_{0}$, while the value used in the SFGX grid is fixed to only $3\zeta_{0}$.
    Such high values imply a considerable increase of heating to the ISM and will in particular affect emission lines from the neutral atomic and molecular gas \citep[e.g.][]{Sternberg1995,Goldsmith2001,Bisbas2021}.
    Here we investigated if the higher CRIR also affects emission from the ionised gas.
    For this purpose we created new \texttt{Cloudy} models using the best parameter set obtained from the model results in Sect. \ref{ssec:ionisedResults} and changed the CRIR from $6\times10^{-16}\,\mathrm{s}^{-1}$ to $10^{-13}\,\mathrm{s}^{-1}$ as reported in \citet{Behrens2022}.
    The \texttt{Cloudy} models were combined according to their weights or covering factors $W_{i}$ as shown in Table \ref{tab:ionisedParamsResults}.
    As expected, emission lines from the ionised gas are hardly affected by a change in the CRIR (see Fig. \ref{fig:CRIR_comparison_ion}).
    The difference in cumulative line flux slightly beyond the ionisation front (i.e. $\xi = 1.25$) is lower than $10\%$ for most of the lines.
    Only Hu\,$\alpha$, $\left[ \ion{Ne}{ii} \right]13\micron$, and $\left[ \ion{S}{iii} \right]33\micron$ have a larger but still small difference of $\lesssim 20\%$.
    Emission lines from the neutral atomic and molecular gas however, are heavily affected by a change of the CRIR.
    Figures \ref{fig:CRIR_comparison_neutral} and \ref{fig:CRIR_comparison_mol} show that the difference in these lines is of the order of one magnitude or even more.

\section{Modelling the ionised and neutral atomic gas}\label{sec:neutralModel}

\subsection{New model grid and sanity checks}\label{ssec:neutral_grid_config}
    To take the higher CRIR and their effect on the neutral atomic (and molecular) gas into account, we created a new sub-grid of SFGX, with an increased CRIR to $10^{-13}\,\mathrm{s}^{-1}$.
    To reduce the computing time for calculating the \texttt{Cloudy} models and for the \texttt{MULTIGRIS} runs, we decreased the parameter space as listed in Table \ref{tab:SFGX+_paramSpace}.
    The resulting new grid contains a total of $10880$ models.
    We combined this new grid with the corresponding sub-grid from SFGX, so that we now have a sampling for the CRIR.
    We will call this new grid SFNX (for \textit{star-forming nucleus with an X-ray source}) in the following\footnote{available at \url{https://doi.org/10.5281/zenodo.8362031}}.

    \begin{table}
        \centering
        \caption{Input parameters of \texttt{Cloudy} models in the new sub-grid of SFGX and parameter space of the original SFGX grid.}
        \begin{tabular}{lll}
            \toprule
            \toprule
            Parameter & SFNX & SFGX \\
            \midrule
            $\log L^{\star}$ [$L_{\odot}$] & $9$                                       & [$7$, $9$] \\
            $L_{X}$ [$L^{\star}$]          & [$0$, $0.001$]                            & [$0$, $0.001$, $0.01$, $0.1$] \\
            $\log T_{X}$ [K]               & [$5$, $6$, $7$]                           & [$5$, $6$, $7$] \\
            $Z$ [$Z_{\odot}$]              & $1$                                       & [$0.01$, $0.02$, $0.1$, \\
                                           &                                           &  $0.33$, $0.5$, $1.0$, $1.2$] \\
            $t$ [Myr]                      & [$1$, $2$, $3$, $4$, $5$, $6$, $8$, $10$] & [$1$, $2$, $3$, $4$, $5$, $6$, $8$, $10$] \\
            $\log n$ [cm$^{-3}$]           & [$0$, $1$, $2$, $3$, $4$]                 & [$0$, $1$, $2$, $3$, $4$] \\
            $\log U$                       & [$-1$, $-2$, $-3$, $-4$]                  & [$0$, $-1$, $-2$, $-3$, $-4$] \\
            $\xi$                          & [$0$, $4$], $\mathrm{step} = 0.25$        & [$0$, $4$], $\mathrm{step} = 0.25$ \\
            \bottomrule
        \end{tabular}
        \label{tab:SFGX+_paramSpace}
    \end{table}

    In Sect. \ref{ssec:ionisedModelCRIReffect} we showed that an increased CRIR only has a minor impact on the line fluxes of ionised gas lines.
    To confirm that the effect is negligible, we proved that \texttt{MULTIGRIS} finds similar parameter sets in SFNX and in SFGX.
    We carried out another run with a two component configuration, using only emission lines originating in the ionised gas as done in Sect. \ref{ssec:ionisedModelConfig}.
    Table \ref{tab:ionisedParamsResults} shows the resulting mean values and confidence intervals for both runs, which are in good agreement.
    The only exception is $L_{X}$, where the results from the new grid are lower but still within uncertainties.
    This is most likely due to a degeneracy effect between $L_{X}$ and $\zeta$ (see Sect. \ref{ssec:CRIRandLX} for a more detailed discussion) and the higher values chosen for $L^{\star} = 10^{9}\,L_{\odot}$ in order to be consistent with the SFGX grid.
    Since the new grid is able to reproduce the results from SFGX, but also takes the much higher CRIR into account, we were now able to investigate solutions for the neutral atomic gas.

\subsection{Results for the ionised and neutral atomic gas}\label{ssec:results_neutral}
    In the next step we added emission lines from the neutral atomic gas with the systematic offsets as determined in Sect. \ref{ssec:ionised_offsets} remaining the same.
    As mentioned earlier, we had not yet included the $\left[ \ion{Fe}{iii} \right]23\micron$ emission line due to the dependence on the dust-to-gas ratio.
    Adding emission lines from the neutral atomic gas such as $\left[ \ion{Fe}{ii} \right]18\micron$ and $\left[ \ion{Fe}{ii} \right]26\micron$ as constraints, we now enable a small systematic offset for the three Fe lines.
    We assume that the line fluxes scale linearly with the iron abundance (within only a small offset), and account for small variations in the dust-to-gas ratio.
    We obtain an offset of $\sim 0.3$ for both configurations, suggesting that the model would under-predict the Fe lines without the scaling.
    
    Since H$_{2}$ rotational transitions arise to a significant amount from the warm neutral atomic gas \citep[e.g.][]{Roussel2007,Togi2016} we also include these lines in these new runs.
    Although we include emission lines from molecular hydrogen, we do not claim that our model is a proper solution for the molecular gas in the centre of NGC\,253.
    
    \begin{table}
        \centering
        \caption{Resulting parameters from an inference using emission lines from the ionised and neutral atomic gas as constraints, for the two different architectures discussed in this study. The model grid used for the inference was SFNX.}
        \begin{tabular}{lrr}
            \toprule
            \toprule
            Parameter & PLaw & 1C2S \\
            \midrule
            $W_{1}$                                      & $1.0_{-0.0}^{+0.0}$       & $0.56_{-0.25}^{+0.26}$ \\
            $W_{2}$                                      & $0.0_{-0.0}^{+0.0}$       & $0.44_{-0.26}^{+0.25}$ \\
            $L_{X}$ [$10^{39}$ erg s$^{-1}$]             & $0.01_{-0.01}^{+0.78}$    & $0.01_{-0.01}^{+0.15}$ \\
            $T_{X}$ [$10^{5}\,\mathrm{K}$]               & $12.18_{-8.68}^{+19.42}$  & $18.33_{-17.17}^{+74.71}$ \\
            $\zeta$ [$10^{-16}\,\mathrm{s}^{-1}$]        & $21.01_{-12.76}^{+45.02}$ & $16.69_{-9.66}^{+38.09}$ \\
            \midrule
            $\alpha_{t}$ [1]                             & $-0.43_{-2.6}^{+2.39}$    & $-$ \\
            $t_{\mathrm{lower}}$ [Myr]                   & $1.79_{-0.35}^{+0.54}$    & $-$ \\
            $t_{\mathrm{upper}}$ [Myr]                   & $2.84_{-0.79}^{+4.96}$    & $-$ \\
            $\bar{t}$ [Myr]                              & $2.33_{-0.56}^{+0.92}$    & $1.94_{-0.46}^{+0.49}$ \\
            \midrule
            $\alpha_{n}$ [1]                             & $-1.69_{-1.12}^{+2.32}$   & $-$ \\
            $n_{\mathrm{lower}}$ / $n_{1}$ [cm$^{-3}$]   & $61_{-59}^{+163}$         & $18_{-14}^{+298}$ \\
            $n_{\mathrm{upper}}$ / $n_{2}$ [cm$^{-3}$]   & $1322_{-1242}^{+5749}$    & $1637_{-1544}^{+7653}$ \\
            $\bar{n}$ [cm$^{-3}$]                        & $228_{-183}^{+516}$       & $730_{-687}^{+3534}$ \\
            \midrule
            $\alpha_{U}$ [1]                             & $-0.83_{-0.5}^{+0.64}$    & $-$ \\
            $\log U_{\mathrm{lower}}$ / $\log U_{1}$ [1] & $-3.72_{-0.23}^{+0.58}$   & $-3.03_{-0.44}^{+0.47}$ \\
            $\log U_{\mathrm{upper}}$ / $\log U_{2}$ [1] & $-2.91_{-0.4}^{+0.39}$    & $-3.38_{-0.46}^{+0.70}$ \\
            $\log \bar{U}$ [1]                           & $-3.22_{-0.29}^{+0.35}$   & $-3.18_{-0.45}^{+0.57}$ \\
            \midrule
            $\alpha_{\xi}$ [1]                           & $-0.96_{-2.1}^{+1.81}$    & $-$ \\
            $\xi_{\mathrm{lower}}$ / $\xi_{1}$ [1]       & $1.2_{-0.63}^{+0.67}$     & $1.27_{-0.45}^{+0.85}$ \\
            $\xi_{\mathrm{upper}}$ / $\xi_{2}$ [1]       & $3.05_{-0.96}^{+0.61}$    & $3.72_{-1.08}^{+0.27}$ \\
            $\bar{\xi}$ [1]                              & $2.18_{-0.7}^{+0.53}$     & $2.35_{-0.73}^{+0.59}$ \\
            \bottomrule                 
        \end{tabular}
        \label{tab:resultParams_neutral+ionised}
    \end{table}

    \begin{figure*}
        \centering
        \includegraphics[width=17cm]{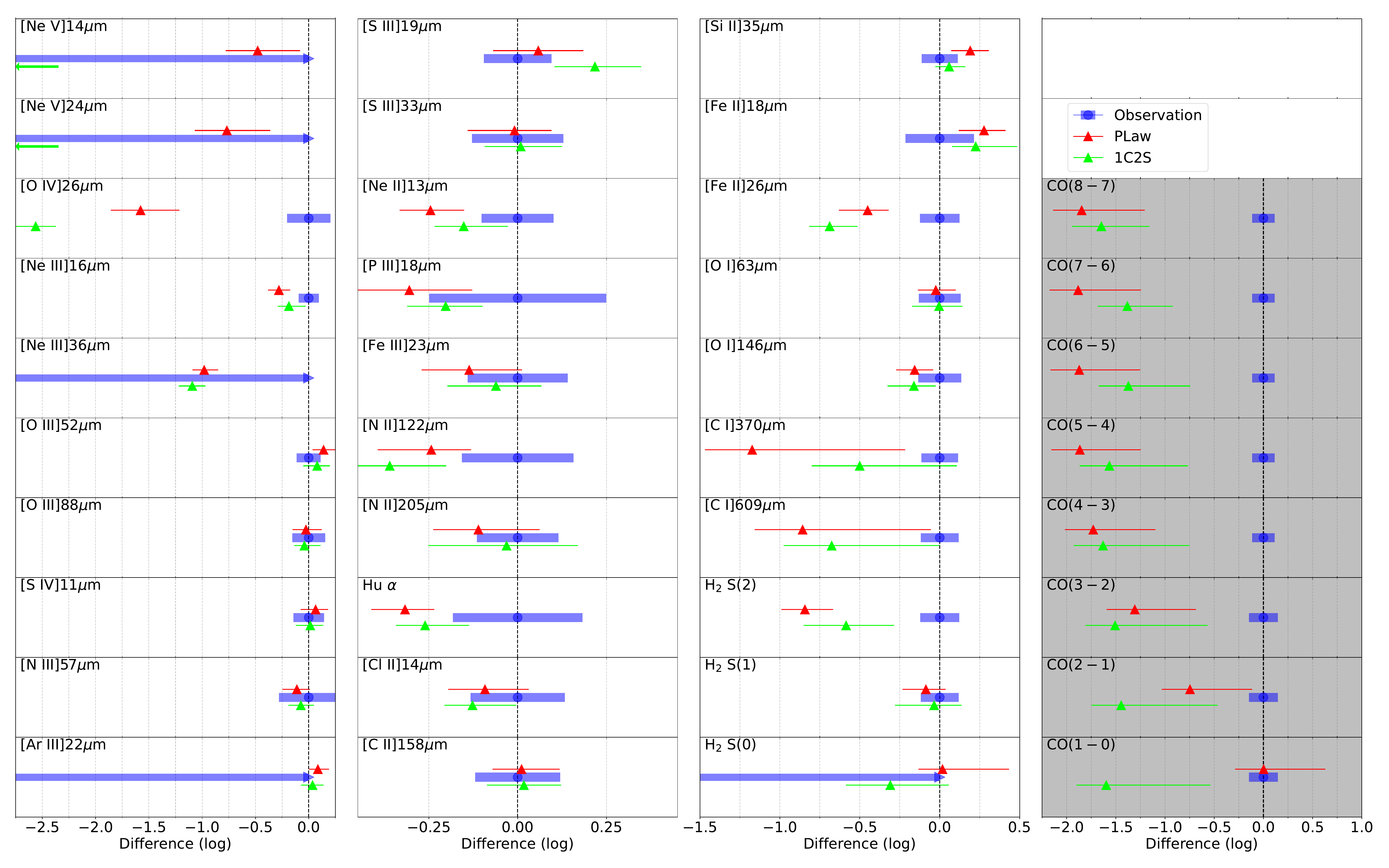}
        \caption{Same as Fig. \ref{fig:ionised_lineFluxComp_PLaw_1C2S}. Emission lines from the ionised and neutral atomic gas are used as constraints (white background) with CO emission lines predicted by \texttt{MULTIGRIS} (grey background).}
        \label{fig:all_lineFluxComp_PLaw_1C2S}
    \end{figure*}
    
    Most of the parameters obtained for the ionised gas model (Sect. \ref{ssec:ionisedResults} and Table \ref{tab:ionisedParamsResults}) are consistent with the model of the ionised and neutral atomic gas as shown in Table \ref{tab:resultParams_neutral+ionised}.
    This shows that the approach to start with only ionised gas lines and determining systematic offsets between instruments (Sect. \ref{ssec:ionised_offsets}) is valid.
    Generally, the preferred model in the PLaw architecture has a high abundance of low density, low ionisation parameter and low depth (i.e. diffuse gas) clouds.
    All three parameters ($n$, $U$, and $\xi$) have a negative slope of the power-law, meaning that higher density clouds with larger depths and higher ionisation parameter are less abundant.
    In the 1C2S architecture, this is reflected by a smaller covering factor $W_{2}$ for the second component.
    However, some parameters show differences compared to the results from the ionised gas.
    Obviously, the cut parameter $\xi$ is larger than in the model for the ionised gas.
    To reproduce emission lines from the neutral atomic gas, the model has to reach values beyond the ionisation front ($\xi = 1$) and even the H/H$_{2}$ photo-dissociation front ($\xi=2$). 
    Since we do not include CO or other emission lines from the molecular core, larger depths ($\xi \geqslant 3$) are not needed nor found.
    The obtained average $\xi \approx 2.3$ (cf. Table \ref{tab:cutParamDefinition}) translates into an $A_{V}$ of $4.5$ which is in good agreement with our results in \citetalias{Beck2022} (see also Fig. \ref{fig:KDE_Av_xi}).
    Furthermore, the density ---in particular the upper limit $n_{\mathrm{upper}}$ for the PLaw architecture and the density of the second component $n_{2}$ for the 1C2S architecture--- is significantly higher than in the ionised gas model.
    This is expected, since the model needs to account for higher critical densities of many of the emission lines from the warm neutral gas such as $\left[ \ion{O}{i} \right]63\micron$ or $\left[ \ion{Si}{ii} \right]35\micron$.
    
    The additional parameter introduced for the CRIR, $\zeta$, is $\sim 2 \times 10^{-15}\,\mathrm{s}^{-1}$.
    This is significantly lower than predicted by \citet{Behrens2022} ($\sim10^{-13}\,\mathrm{s}^{-1}$), which is expected for two reasons.
    First, the region observed is much larger in our case ($40\arcs$ compared to $1.6\arcs$ in \citealt{Behrens2022}), so the locally increased CRIR from \citet{Behrens2022} is smeared out within our observing beam.
    Second, although many lines are sensitive to the CRIR as shown in Sect. \ref{ssec:ionisedModelCRIReffect}, higher fluxes for these lines can be modelled for instance by a higher density, a stronger radiation field and/or a higher X-ray luminosity (see Sect. \ref{ssec:CRIRandLX}).
    It is possible that there is some degeneracy between these parameters, which may also be reflected by the large uncertainties.
    
    Some of the emission lines are not reproduced by any of the architectures (e.g. $\left[ \ion{N}{ii} \right]122\micron$ or $\left[ \ion{Fe}{ii} \right]26\micron$), resulting in lower $pn\sigma$ values, now with $p3\sigma$ of $85\%$ compared to $\geqslant 95\%$.
    See Fig. \ref{fig:all_lineFluxComp_PLaw_1C2S} for a comparison of observed and modelled line fluxes for both architectures.
    The $pn\sigma$ values are, however, still high enough to obtain reliable inference of further (secondary) parameters from this model (Sect. \ref{sec:secondaryParams}).
    
\section{Discussion}\label{sec:discussion}

\subsection{Cosmic ray ionisation rate and X-ray luminosity}\label{ssec:CRIRandLX}
    Table \ref{tab:ionisedParamsResults} shows that the X-ray luminosity drops significantly from $\sim5\times10^{39}\,\mathrm{erg\,s}^{-1}$ to $\sim 10^{37}\,\mathrm{erg\,s}^{-1}$, once the CRIR is increased.
    This somewhat contradicts our finding that X-rays are needed to reproduce the highest ionised emission lines in our sample.
    Hence, we performed another \texttt{MULTIGRIS} run, forcing a higher X-ray luminosity of $L_{X} = 5\times10^{39}\,\mathrm{erg\,s}^{-1}$.
    The resulting parameter sets for both architectures (PLaw and 1C2S) are similar, with the exception of the CRIR, which drops by $\sim25\%$ in the higher X-ray runs, consistent with findings from \citet{Lebouteiller2017}.
    Cosmic rays affect mostly the neutral atomic and molecular gas, while the ionised gas remains unchanged.
    X-rays, however, affect the ionised gas ---and high ionisation states in particular--- but also heat the neutral atomic gas, which creates a degeneracy between the luminosity of the X-ray source and the CRIR.
    Such difficulties between the discrimination of X-rays and cosmic rays has already been shown, for instance by \citet{Meijerink2006}.
    Observations of the CO$^{+}$ molecule could help to break this degeneracy, as done in M82 \citep{Spaans2007}.
    In conclusion, this implies that both, the CRIR and X-ray luminosity obtained in this study (Tables \ref{tab:ionisedParamsResults} and \ref{tab:resultParams_neutral+ionised}) should rather be considered as upper limits.
    
\subsection{Predictions for the molecular gas}\label{ssec:predict_molGas}
    From the PDFs of the model parameters, \texttt{MULTIGRIS} is able to predict the luminosity of other emission lines.
    We predicted line fluxes for several CO emission lines ($J = 1 \rightarrow 0$ to $J = 8 \rightarrow 7$) that arise from the molecular gas.
    We compared the predicted line fluxes with those observed by \textit{Herschel}/PACS and \textit{Herschel}/SPIRE \citep[see][]{PerezBeaupuits2018} but did not use them as constraints.
    Figure \ref{fig:all_lineFluxComp_PLaw_1C2S} compares the observed and predicted CO line fluxes and the corresponding uncertainties in column 4 (grey background).
    With the exception of CO($1-0$) in the 1C2S architecture, all CO emission lines are under-predicted by up to two orders of magnitude.
    The model clearly fails to account for the molecular gas.

    We investigated potential causes for the under-prediction of the CO emission.
    However, for a more detailed model of the molecular gas we refer to a forthcoming paper.
    Due to the increased CRIR, the ISM is much warmer and the H/H$_{2}$ and C/CO photo-dissociation front are shifted to larger depths with increasing $\zeta$.
    Since the \texttt{Cloudy} models, however, stop at a fixed $A_{V} = 10\,\mathrm{mag}$, models with a higher CRIR contain less H$_{2}$ and therefore also less CO.
    This, in turn, results in significantly lower CO line fluxes simply due to the low abundance of CO in the whole \texttt{Cloudy} model.
    To overcome the low CO abundance, we investigated which $A_{V}$ is needed to predict the observed CO line fluxes, by running new \texttt{Cloudy} models with parameters from Table \ref{tab:resultParams_neutral+ionised} but with an arbitrary high $A_{V}$ of $30\,\mathrm{mag}$.
    In fact, such a model would be able to reproduce the CO line fluxes as observed by \textit{Herschel}/SPIRE and JCMT, but the increase of depth obviously affects other lines in the model as well.
    In particular emission lines like the $\left[ \ion{C}{i} \right]$ and $\left[ \ion{O}{i} \right]$ lines originate partly in the molecular gas and become brighter with larger depth.
    See also Fig. \ref{fig:CRIR_comparison_neutral} showing that the cumulative fluxes of these lines are growing even at larger $\xi$ values.
    Furthermore, we calculated the total molecular hydrogen mass $M(\mathrm{H}_{2})$ from these high $A_{V}$ models (see Sect. \ref{sec:secondaryParams}), which is two orders of magnitude larger than typical hydrogen masses in the nuclei of galaxies.
    To conclude, simply increasing the depth of the models cannot overcome the under-predicted CO emission for which other emission lines would be affected as well as an absurdly high molecular gas mass would be obtained.
    Furthermore, it would be in contradiction with our results for an $A_{V}$ of $4.35$ from the SED fit in \citetalias{Beck2022}.

    Besides a higher $A_{V}$, shocks could play a role in the excitation of CO emission \citep[e.g.][]{Lesaffre2013,Pon2016,Kamenetzky2018}.
    Furthermore, \citet{Hao2009} showed that shocks can possibly excite $\left[ \ion{O}{iv} \right]26\micron$ emission, which is significantly under-predicted in our models as well.
    Another mechanism not taken into account is the time-dependence of the chemistry.
    \texttt{Cloudy} assumes a static chemistry, however, simulations have shown that a time-dependent chemistry (in particular in nuclei of active galaxies) can result in several orders of magnitude brighter CO line fluxes \citep{Meijerink2013}.


\subsection{Effects of the initial mass function}\label{sec:effect_IMF}
    To determine the emitted spectrum of an interstellar cloud, \texttt{Cloudy} requires an input spectrum, in our case the spectrum of a starburst cluster.
    The distribution of the overall mass among the stars within the cluster (the so-called \textit{initial-mass-function}, IMF) can have a significant effect on the output spectrum of the cluster -- and hence also on the physical conditions and spectrum of the illuminated cloud.
    However, the exact shape of the IMF and whether it universal, is still uncertain and debated \citep[e.g.][ and references therein]{Hopkins2018}.
    We analysed the impact of a different IMF, by creating a new input spectrum -- to eliminate potential uncertainties by using different codes (and therefore different assumptions such as evolutionary tracks, etc.), we used BPASS to predict the stellar emission.
    While the other parameters remain the same \citep{Eldridge2017,Ramambason2022}, we change the IMF from a Kroupa-like IMF \citep{Kroupa1993} to a Salpeter IMF \citep{Salpeter1955}.
    Thereafter, we ran two \texttt{Cloudy} simulations with the parameters from Table \ref{tab:resultParams_neutral+ionised}, one using the spectrum from a Kroupa IMF, and one using the spectrum from a Salpeter IMF.

    \begin{figure}
        \centering
        \resizebox{\hsize}{!}{\includegraphics[scale=1]{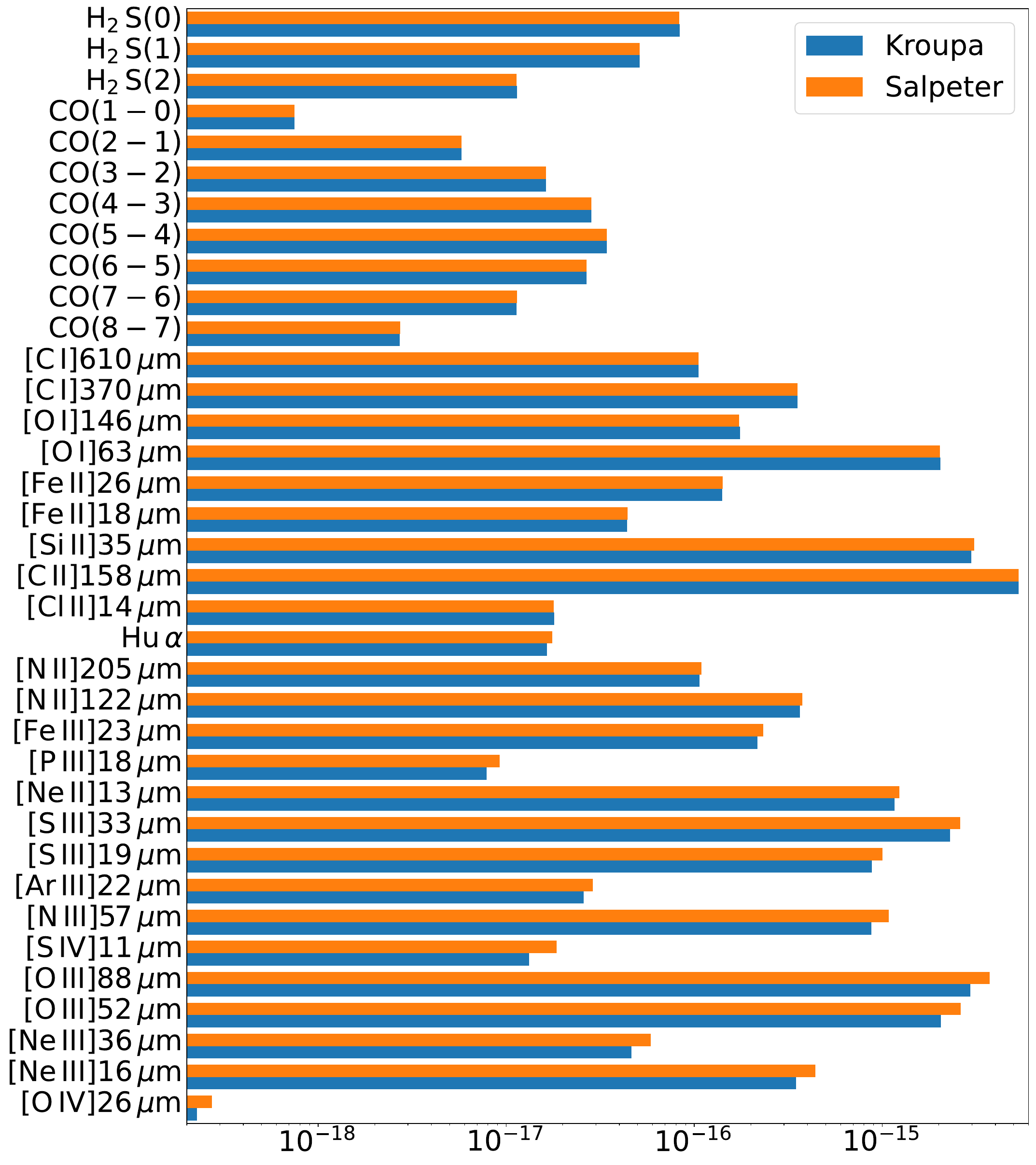}}
        \caption{Comparison of predicted line fluxes from \texttt{Cloudy} runs, using a Kroupa (blue) and Salpeter (orange) IMF input spectrum, respectively. For representation reasons the $\left[ \ion{Ne}{v} \right]$ emission lines are not shown but are both of the order of $10^{-28}\,\mathrm{W\,m}^{-2}$, i.e. within the upper limits of our observations.}
        \label{fig:comp_IMF_BPASS_SB99}
    \end{figure}

    Figure \ref{fig:comp_IMF_BPASS_SB99} shows the resulting line fluxes from these \texttt{Cloudy} simulations.
    The line fluxes from the two models are generally in good agreement, in particular emission lines originating in the neutral atomic (e.g. $\left[ \ion{O}{i} \right]$ and $\left[ \ion{C}{i} \right]$) and molecular gas (CO and H$_{2}$), where they agree within $1\%$.
    Emission lines, that to some fraction come from the ionised gas, are slightly more affected.
    The $\left[ \ion{Fe}{ii} \right]$ lines, $\left[ \ion{C}{ii} \right]158\micron$, and $\left[ \ion{Si}{ii} \right]35\micron$ differ by $\sim 5\%$.
    The largest deviations occur for emission lines that purely arise from the ionised gas, with the Kroupa IMF yielding fainter line fluxes than the Salpeter IMF in almost all cases.
    For instance, the two $\left[ \ion{S}{iii} \right]$ lines are brighter by $\sim10\%$, the two $\left[ \ion{Ne}{iii} \right]$ lines even by $\sim20\%$.
    However, the ratio of two emission lines from the triplet of a species (e.g. $\left[ \ion{N}{ii} \right]$, $\left[ \ion{S}{iii} \right]$, $\left[ \ion{O}{iii} \right]$) are constant for one IMF.
    This indicates that the reason for the differences is primarily due to different abundances of the respective ions.
    Indeed, the abundance of higher ionic states (which is one output of \texttt{Cloudy}) are $5-10\%$ higher for the Kroupa IMF at depths lower than the ionisation front.

    We conclude that the particular shape of the chosen IMF can have a significant impact on the resulting line fluxes, in particular those from emission lines originating in the ionised gas.
    However, by selecting a state-of-the-art IMF and model BPASS, we assume that these effects are minor and that our results are valid.
    

\subsection{Secondary parameters obtained from the model}\label{sec:secondaryParams}
    In addition to predicting line fluxes of other emission lines, \texttt{MULTIGRIS} also allows us to predict secondary parameters in post-processing runs.
    These secondary parameters must be stored in a post-processing grid, which is associated with the main grid, in our case SFGX and SFNX, respectively.
    The values listed throughout this section are results from the PLaw architecture.
    However, similar to the results of the primary parameters in Table \ref{tab:resultParams_neutral+ionised}, the post-processing parameters are comparable and agree within uncertainties.
    Assuming a spherical geometry, masses of dust and the different hydrogen phases were calculated from the \texttt{Cloudy} output.
    Using our final model (see Table \ref{tab:resultParams_neutral+ionised}) we are then able to estimate the mass of ionised and neutral atomic hydrogen in the centre of NGC\,253.
    In principle, we can also obtain a mass for the molecular hydrogen.
    However, due to the limited validity of our model regarding the molecular gas, results for this phase have to be taken with caution, as noted by the large uncertainties associated with these parameters.
    The resulting masses are shown in Table \ref{tab:postProcess_params}.
    Both, the mass for the ionised hydrogen and for neutral atomic hydrogen are in good agreement with estimates from KAO observations \citep{Carral1994}, who obtained $1\times10^{6}\,M_{\odot}$ and $5\times10^{6}\,M_{\odot}$ from the \citet{Tielens1985} PDR models for the ionised and neutral atomic hydrogen, respectively.
    
    \begin{table}
        \centering
        \caption{Resulting post-processing parameters for the nucleus of NGC\,253.}
        \begin{tabular}{lr}
            \toprule
            \toprule
            Parameter & Value \\
            \midrule
            $M(\mathrm{H}^{+})$ [$10^{6}\,M_{\odot}$] & $3.81_{-1.15}^{+6.81}$ \\
            $M(\mathrm{H}^{0})$ [$10^{6}\,M_{\odot}$] &	$9.14_{-1.83}^{+10.02}$ \\
            $M(\mathrm{H}_{2})$ [$10^{7}\,M_{\odot}$] & $19.58_{-13.29}^{+46.68}$ \\
            $M(\mathrm{dust})$ [$10^{5}\,M_{\odot}$]  & $1.80_{-1.10}^{+3.30}$ \\
            \midrule
            $\left[ \ion{C}{ii} \right](\mathrm{H}^{+})$ & $26\%$ \\
            $\left[ \ion{C}{ii} \right](\mathrm{H}^{0})$ & $35\%$ \\
            $\left[ \ion{C}{ii} \right](\mathrm{H}_{2})$ & $39\%$ \\
            \midrule 
            $F(\mathrm{H}\alpha)$ [$10^{-14}\,\mathrm{W\,m}^{-2}$]& $7.84_{-1.66}^{+9.57}$ \\
            \bottomrule
        \end{tabular}
        \label{tab:postProcess_params}
    \end{table}

    We can also estimate the fraction of $\left[ \ion{C}{ii} \right]158\micron$ associated with the ionised, neutral atomic, and molecular gas.
    Recently,  $\left[ \ion{C}{ii} \right]$ has been increasingly considered as a probe for the CO dark gas or even the total molecular gas mass \citep[e.g.][and references within]{Madden2020}.
    It is necessary, however, to correct for emission coming from \ion{H}{ii} regions that do not contain any H$_{2}$.
    In the case of NGC\,253 with solar metallicity, the fraction of $\left[ \ion{C}{ii} \right]158\micron$ from the ionised gas phase is $\sim 26\%$ (see Table \ref{tab:postProcess_params}).
    This is somewhat lower than the findings in \citet{Cormier2019} (using their Eq. (2) yields $45\%$ of $\left[ \ion{C}{ii} \right]158\micron$ from the ionised gas), which could be due to different CRIR assumed in our study or the fact that solar metallicities are not covered in \citet{Cormier2019}.

\subsection{Star-formation rates}\label{sec:SFRs}
    In Sect. \ref{ssec:ionisedResults} we showed that the potential AGN in the centre of NGC\,253 has little effect on its environment as probed by the observed lines.
    What really drives the heating in the nucleus seems to be the star-formation activity, usually quantified by the star-formation rate (SFR).
    Within the last few decades, a number of tracers for the SFR have been proposed, either from theoretical considerations or empirical calibrations \citep[see][for a review]{Kennicutt2012}.
    Each of these tracers has its (dis-)advantages, in particular regarding where the respective probe arises, and if it is affected by extinction.
    In this section we will compare three different SFR-tracers, namely the H$\alpha$, $\left[\ion{C}{ii}\right]158\micron$, and $L_{\mathrm{TIR}}$ luminosities.
    Throughout we assume a distance of $3.5\,\mathrm{Mpc}$ \citep{Rekola2005} to convert fluxes to luminosities.

    One of the most important tracers of the SFR is the luminosity of the H$\alpha$ line, since it is easily observable with optical telescopes and is directly emitted by young stars.
    From our latest model presented in Sect. \ref{sec:neutralModel}, \texttt{MULTIGRIS} is able to infer the intrinsic (extinction-free) luminosity of this emission line (Table \ref{tab:postProcess_params}).
    Using the empirical calibration from \citet{Calzetti2007}, which is also tested on smaller, $\sim100\,\mathrm{pc}$ scales \citep{Pessa2021,Belfiore2023}, we obtain $\mathrm{SFR} = 2.3^{+3.1}_{-0.4}\,M_{\odot}\,\mathrm{yr}^{-1}$.
    One disadvantage of the H$\alpha$ emission line is that that it  that it typically suffers from extinction in extra-galactic sources, which is why we corrected the line flux from Table \ref{tab:postProcess_params} assuming a mixed extinction model and the optical depth of $4.5\,\mathrm{mag}$ as determined in Sect. \ref{ssec:results_neutral}.

    Another frequently employed probe for the SFR is the total infrared luminosity $L_{\mathrm{TIR}}$.
    It assumes that in thermal equilibrium, most of the stellar radiation is reprocessed by dust and radiated within the infrared spectral range (i.e. between $3\micron$ and $1000\micron$).
    We determined the $L_{\mathrm{TIR}}$ already in \citetalias{Beck2022} in two different ways, confirmed by our \texttt{MULTIGRIS} approach.
    The calibration shown in \citet{Kennicutt1998} yields $\mathrm{SFR} = 1.6\pm 0.4\,M_{\odot}\,\mathrm{yr}^{-1}$, which is in good agreement with the SFR determined from H$\alpha$. 

    \citet{Stacey1991} proposed the $\left[\ion{C}{ii}\right]158\micron$ emission line as a probe for the SFR.
    Since we directly measured this quantity and did not calculate or predict it, we believe that this is the most reliable measurement of the SFR in this study -- not taking any calibration uncertainties into account.
    Using the more recent calibration from \citet{HerreraCamus2018}, we obtain $\mathrm{SFR} = 1.7\pm 0.5 \,M_{\odot}\,\mathrm{yr}^{-1}$.

    All the SFRs we obtained from the different probes are in good agreement with each other.
    They are also comparable with the results from ISO observations in \citet{Radovich2001}, who obtained $\mathrm{SFR} = 2.1\,M_{\odot}\,\mathrm{yr}^{-1}$ for the nucleus, however, with a larger beam of the ISO telescope compared to our observations.
    The larger observing beam could result in contamination of infrared emission from the disk and therefore in a slightly higher SFR.
    Yet, the result from \citet{Radovich2001} is within the uncertainties of our solution.

\subsection{Comments and caveats of the model grids}
    Although the model grids SFGX and SFNX that we use cover a wide range of ISM conditions, we note that some mechanisms are not taken into account which may affect the model results, and in particular might lead to the under-prediction of the CO emission lines.
    These aspects shall be shortly discussed and will be further investigated in future studies.

    According to \citet{Sanchez2020} or \citet{Sanchez2021}, one potential major contribution to the radiation field can be post-asymptotic-giant-branch (AGB) stars. 
    These objects typically have a hard but weak ionising effect on the ISM, which could cascade down into the CO emitting regions.
    AGB stars are included in the BPASS models, however, the post-AGB phase is not well understood and needs improvements in the model \citep{Eldridge2017}.
    While the works of \citet{Sanchez2020,Sanchez2021} show that ionisation occurs on local scales, the limited spatial resolution of our observations in this work leads to the fact that the ionising sources are not resolved.

    Another mechanism not considered in the model grids are shocks, for instance from supernovae or galactic outflows.
    Such outflows have been observed in CO, H$\alpha$, and X-ray emission \citep{Bolatto2013}.
    They are well known to contribute to ISM heating and therefore boost emission lines.

    Lastly, the relative abundances of elements varies not only within galaxies but even when comparing \ion{H}{ii} regions within one galaxy \citep{GarciaRojas2007}.
    The model grids take into account that the abundance, for example of N and C, varies with the O/H ratio.
    However, it might also be the case that the N and C relative abundance changes as well.
    A deviation in the relative abundances of elements would directly lead to a change of the relative line fluxes of different species.

\section{Summary}
    In this study we used a combined set of $30$ emission lines from SOFIA, \textit{Herschel}, and \textit{Spitzer} observations of the nuclear region of the starburst galaxy NGC\,253 to investigate the physical conditions of the ISM on a $100\,\mathrm{pc}$ scale.
    Using \texttt{MULTIGRIS}, a Bayesian code to probabilistically investigate a set of \texttt{Cloudy} models, we constrained parameters of the ionised and neutral atomic gas.
    After eliminating systematic offsets between the different telescopes, instruments, and modules, we first determined parameters of the ionised gas.
    The created model was able to reproduce all observed emission lines with the exception of $\left[ \ion{O}{iv} \right]26\micron$.
    We found that the metallicity and density that we calculated from analytic prescriptions in \citetalias{Beck2022} are in good agreement with our probabilistic results.
    Furthermore, we inferred that the hypothetical AGN in the nucleus of NGC\,253 has a minor impact on the heating of the ISM, with luminosities $\lesssim 6 \times 10^{39}\,\mathrm{erg\,s}^{-1}$ or $\lesssim 1.5 \times 10^{6}\,L_{\odot}$.
    After modelling the ionised gas, we added emission lines originating in the neutral atomic gas, with an increased CRIR as shown by \citet{Behrens2022}.
    We showed that the higher CRIR has little to no effect on the results for the ionised gas.
    Again, the extended model is able to reproduce most ($24$ of $30$) of the emission lines, and the obtained optical depth is in good agreement with our results from \citetalias{Beck2022}.
    However, the model fails to reproduce most of the CO emission -- we refer to a future study to further investigate the molecular gas properties.
    From our model we were able to calculate gas and dust masses, and determined the fraction of $\left[ \ion{C}{ii} \right]158\micron$ emission from the different phases (Table \ref{tab:postProcess_params}). 
    Since the nuclear starburst seems to have the major heating impact on the ISM, we calculated the star-formation rate from different tracers which are all in good agreement ($0.6 - 1.7\,M_{\odot}\,\mathrm{yr}^{-1}$).    
    
\begin{acknowledgements}
    We thank the referee, S. S\'{a}nchez, for comments which improved the clarity of the paper.
    The authors acknowledge support by the state of Baden-Württemberg through bwHPC.
    LR gratefully acknowledges funding from the DFG through an Emmy Noether Research Group (grant number CH2137/1-1).
\end{acknowledgements}

\bibliography{Literature}

   \bibliographystyle{aa} 

\appendix

\section{Relation between $A_{V}$ and $\xi$}
\begin{figure}
        \centering
        \resizebox{\hsize}{!}{\includegraphics[scale=1]{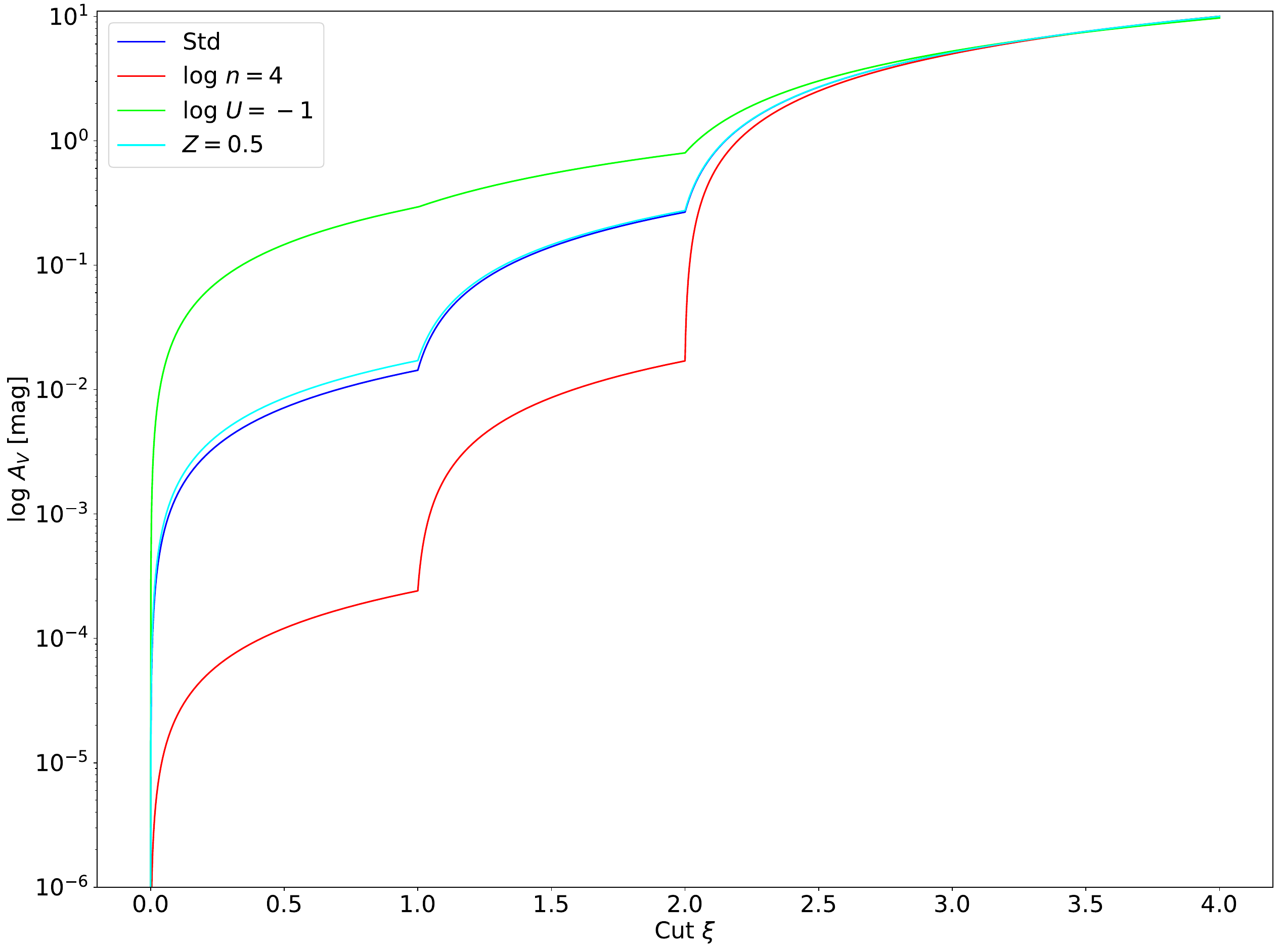}}
        \caption{Relation between $A_{V}$ and $\xi$ from \texttt{Cloudy} models with different ISM parameters. The \textit{Std} model results from a parameter set with $\log n = 2\,\mathrm{cm}^{-1}$, $\log U = -3$, and $Z=1\,Z_{\odot}$. The legend denotes which parameters was changed.}
        \label{fig:Av_vs_cut}
    \end{figure}

    The definition of the cut parameter $\xi$ (Table \ref{tab:cutParamDefinition}) is linked to the density of the ionisation states of hydrogen.
    Therefore, the optical (and physical) depth of the $\xi$ values change with the ambient conditions, such as the density $n$, ionisation parameter $U$, or metallicity $Z$.
    Specifically, there is no equation describing the relation between $\xi$ and $A_{V}$.
    Figure \ref{fig:Av_vs_cut} illustrates the relation between these to parameters, with changing ambient conditions from \texttt{Cloudy} models.
    The \textit{Std} model (blue) shows the resulting $A_{V}$ vs $\xi$ relation, with a parameter set of $\log n = 2\,\mathrm{cm}^{-3}$, $\log U = -3$, and $Z=1\,Z_{\odot}$, which is close the final parameter set (Sect. \ref{ssec:results_neutral}).
    Red, green, and cyan lines show the relation with changing density ($\log n = 4\,\mathrm{cm}^{-3}$), ionisation parameter $\log U = -1$, and metallicity $Z=0.5\,Z_{\odot}$, respectively.

\begin{figure}
        \centering
        \resizebox{\hsize}{!}{\includegraphics[scale=1]{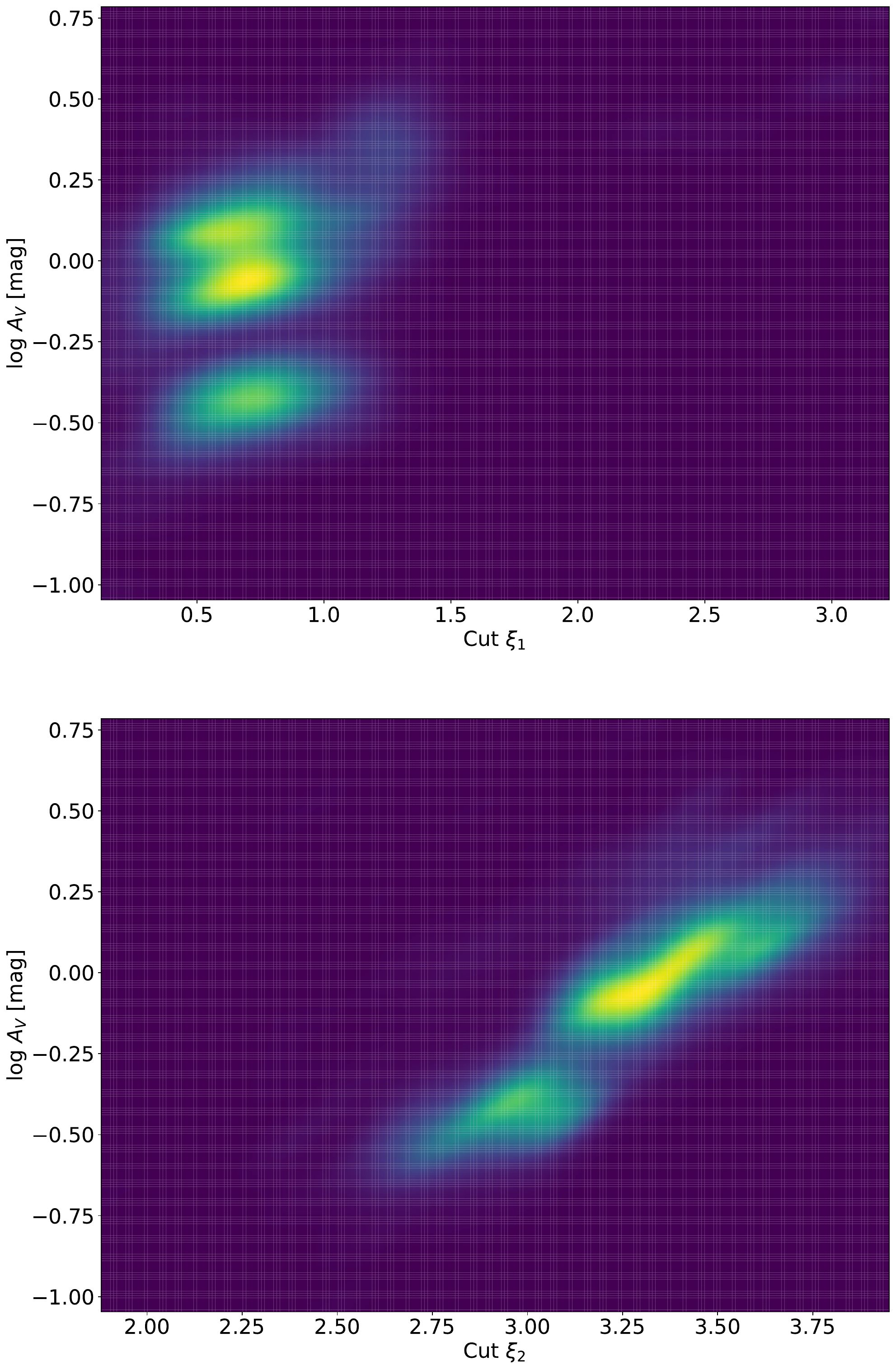}}
        \caption{Kernel density estimate (KDE) plots showing the relation between $A_{V}$ and $\xi$ from the \texttt{MULTIGRIS} runs as described in Sect. \ref{sec:neutralModel} for a 1C2S configuration.}
        \label{fig:KDE_Av_xi}
    \end{figure}

    Figure \ref{fig:KDE_Av_xi} shows the kernel-density-estimate plots from the \texttt{MULTIGRIS} run of a 1C2S configuration as described in Sect. \ref{sec:neutralModel}.
    This illustrates which values for $\xi_{1}$ and $\xi_{2}$ are a likely solution, and to which $A_{V}$ this corresponds.
    The spread in the $A_{V}$ for similar $\xi$, in particular in the top panel corresponds to the different physical conditions (e.g. different metallicities or densities).
    
\section{Effects of cosmic rays on emission lines from the ionised gas}
    For completeness reasons we show the emissivity and cumulative line flux of lines from the ionised gas for two different CRIR $\zeta$ (see Sect. \ref{ssec:ionisedModelCRIReffect}).
    Since cosmic rays do not dominate the heating in \ion{H}{ii} regions, the effect of an increased CRIR ---in particular before the ionisation front ($\xi = 1$)--- is negligible.
    
    \begin{figure*}
        \centering
        \includegraphics[width=17cm]{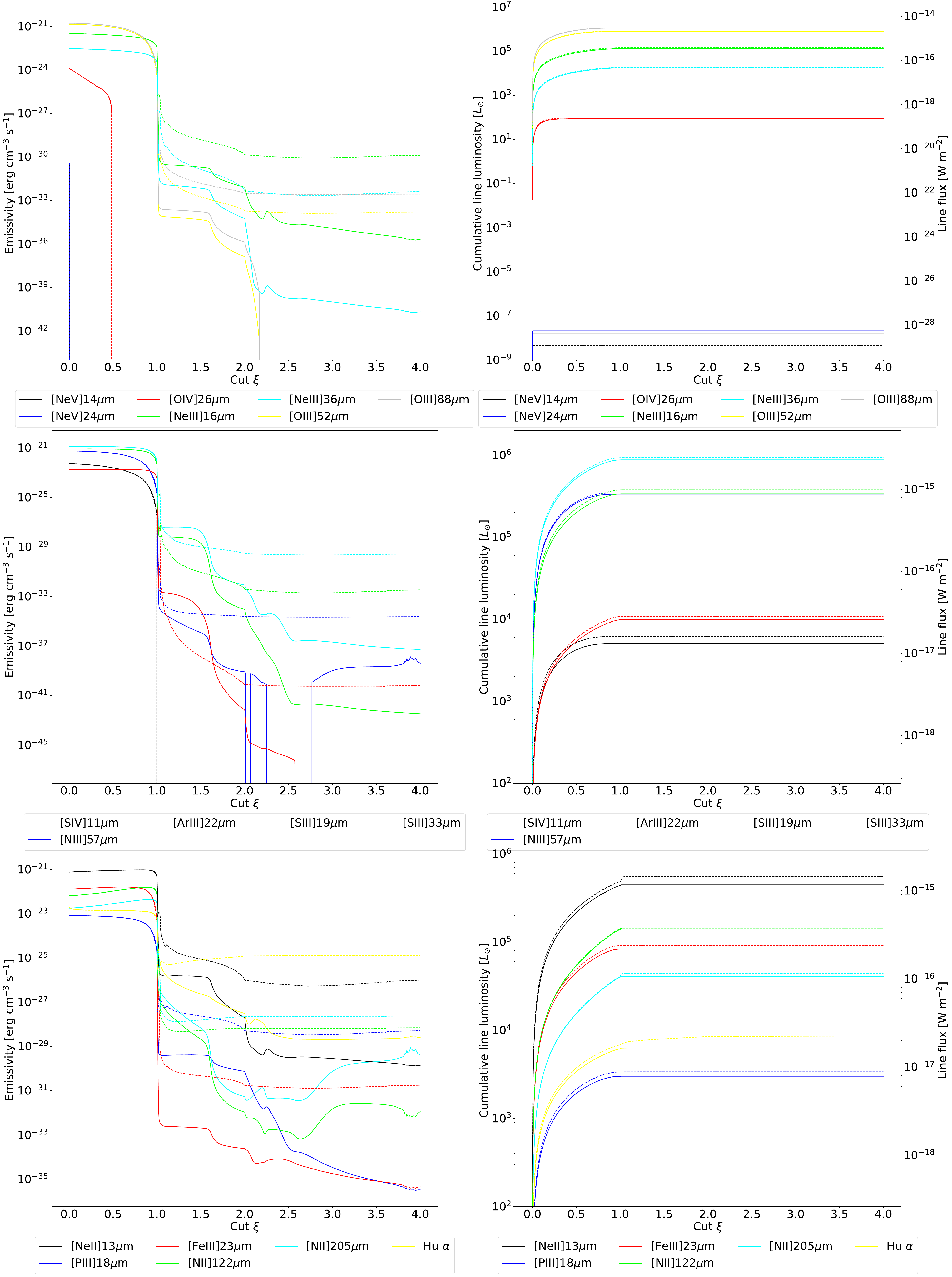}
        \caption{Emissivity (\textit{left}) and cumulative line flux (\textit{right}) over the cut parameter $\xi$ for emission lines from the ionised gas. Solid lines show models with a low CRIR ($6\times10^{-16}\,\mathrm{s}^{-1}$), dashed lines are from models with a high CRIR ($10^{-13}\,\mathrm{s}^{-1}$).}
        \label{fig:CRIR_comparison_ion}
    \end{figure*}    

\end{document}